\documentclass[12pt]{article}

\usepackage[top=30mm, bottom=40mm, left=25mm, right=25mm]{geometry}
\usepackage{microtype}
\usepackage{amssymb}
\usepackage{amsmath}
\usepackage{mathrsfs}
\usepackage{fancyhdr}
\usepackage{marvosym}
\usepackage[mathcal]{euscript}
\usepackage{indentfirst}
\usepackage{forloop}
\usepackage{etoolbox}
\usepackage[symbol=\!]{footnotebackref} 
\usepackage{lmodern}
\usepackage{slantsc}
\usepackage[noadjust]{cite} 
\usepackage{letltxmacro}

\newcommand{\tamusymbol}{*}
\newcommand{\berkeleysymbol}{\dagger}
\newcommand{\albanysymbol}{\ddagger}
\renewcommand{\d}{{\mathrm d}}

\newcommand{\D}{\mathcal{D}}

\newcommand{\Q}{\mathcal{Q}}

\newcommand{\R}{\mathbf{R}}
\newcommand{\ad}{{\dot{\alpha}}}

\renewcommand{\epsilon}{\varepsilon}

\makeatletter
\newcommand{\ceq}{\mathrel{\rlap{\raisebox{0.3ex}{$\m@th\cdot$}}\raisebox{-0.3ex}{$\m@th\cdot$}}=}
\newcommand{\eqc}{=\mathrel{\rlap{\raisebox{0.3ex}{$\m@th\cdot$}}\raisebox{-0.3ex}{$\m@th\cdot$}}}
\makeatother
\newtoggle{eprint}
\newcommand{\inspire}[1]{[\href{https://inspirehep.net/record/#1}{\texttt{inSPIRE:#1}}]}

\newcommand{\journal}[4]{\href{#2}{#1} \textbf{#3} #4}
\newcommand{\key}{}
\newcounter{zero}
\makeatletter
\renewenvironment{thebibliography}[1]
   {\section*{\refname}%
       \@mkboth{\MakeUppercase\refname}{\MakeUppercase\refname}%
       \list{\@biblabel{\hyperlink{\key}{\@arabic\c@enumiv}}}%
       {\settowidth\labelwidth{\@biblabel{#1}}%
           \leftmargin\labelwidth
           \advance\leftmargin\labelsep
           \@openbib@code
           \usecounter{enumiv}%
           \let\p@enumiv\@empty
           \renewcommand\theenumiv{\@arabic\c@enumiv}}%
       \sloppy
       \clubpenalty4000
       \@clubpenalty \clubpenalty
       \widowpenalty4000%
       \sfcode`\.\@m}
   {\def\@noitemerr
       {\@latex@warning{Empty `thebibliography' environment}}%
       \endlist}
\makeatother
\newcommand{\targetlist}[1]{\forcsvlist\target{#1}}
\newcommand{\target}[1]{\ucite{#1}\raisebox{\ht\strutbox}{\hypertarget{#1\arabic{#1ctr}}{}}}
\LetLtxMacro{\oldcite}{\cite}
\renewcommand{\cite}[1]{\targetlist{#1}\protect\oldcite{#1}}
\newcommand{\addocc}[1]{%
   \iftoggle{eprint}{%
   \ifnumcomp{\value{#1ctr}}{>}{0}{%
       \newcounter{#1k}\setcounter{#1k}{1}%
       \addtocounter{#1ctr}{1}
       (\textit{also~cited~at}~\hyperlink{#1\arabic{#1k}}{\textsf{\roman{#1k}}}%
       \forloop{#1k}{2}{\value{#1k} < \value{#1ctr}}{%
           , \hyperlink{#1\arabic{#1k}}{\textsf{\roman{#1k}}}%
       }).%
   }{}}{}}
\newcounter{citecnt}
\newtoks\citetoks
\makeatletter
\newcommand{\ucite}[1]{%
   \@ifundefined{uns@cite#1}
       {\global\newcounter{#1ctr}\setcounter{#1ctr}{0}%
       \refstepcounter{citecnt}\label{citelabel@#1}%
       \expandafter\xdef\csname uns@cite#1\endcsname{\arabic{citecnt}}%
       \toks\z@=\expandafter{\the\citetoks}%
       \toks\tw@=\expandafter\expandafter\expandafter{%
       \csname uns@bibitem#1\endcsname}%
       \edef\@tempcite{\the\toks\z@\the\toks\tw@}%
       \global\citetoks=\expandafter{\@tempcite}%
       }{\refstepcounter{#1ctr}}}
\LetLtxMacro{\oldbibitem}{\bibitem}
\renewcommand{\bibitem}[2]{%
   \@namedef{uns@bibitem#1}{\renewcommand{\key}{#1\arabic{zero}} \oldbibitem{#1} #2 \addocc{#1}}}
\makeatother

\newcommand{\be}{\begin{equation}}
\newcommand{\ee}{\end{equation}}
\newcommand{\bea}{\begin{eqnarray}}
\newcommand{\eea}{\end{eqnarray}}

\newcommand{\al}{\alpha}
\renewcommand{\d}{\delta}
\newcommand{\e}{\epsilon}
\newcommand{\G}{\Gamma}
\newcommand{\g}{\gamma}

\newcommand{\La}{\Lambda}
\newcommand{\la}{\lambda}

\newcommand{\Om}{\Omega}
\newcommand{\om}{\omega}
\newcommand{\s}{\sigma}
\renewcommand{\t}{\theta}

\newcommand{\hlf}{\frac{1}{2}}

\newcommand{\non}{\nonumber}

\newcommand{\p}{\partial}

\newcommand{\w}{\wedge}

\newcommand{\Real}{\operatorname{Re}}

\newcommand{\Tr}{\operatorname{Tr}}

\newcommand{\mcA}{\mathcal{A}}

\newcommand{\mcF}{\mathcal{F}}
\newcommand{\mcL}{\mathcal{L}}

\newcommand{\mcW}{\mathcal{W}}

\newcommand{\hph}[1]{{\hphantom{#1}}}

\usepackage{amsthm} 
\theoremstyle{definition}

\theoremstyle{plain}

\newtheorem*{result*}{Result}
\newtheorem{claim}{Claim}

\toggletrue{eprint} 
\iftoggle{eprint}{
   \newcommand{\srrefcolor}{cyan}
   \newcommand{\srurlcolor}{cyan}
}{
   \newcommand{\srrefcolor}{black}
   \newcommand{\srurlcolor}{black}
}
\numberwithin{equation}{section}
\renewcommand{\arraystretch}{1.3}
\hypersetup{
   colorlinks = true,
   citecolor = \srrefcolor,
   linkcolor = \srrefcolor,
   urlcolor = \srurlcolor,
   bookmarksopen = true,
   bookmarksopenlevel = \maxdimen
}
\title{\textbf{
All Chern-Simons Invariants of 4D, $N=1$ Gauged Superform Hierarchies
} \\[0.5cm]}
\author{\normalsize Katrin Becker,\!$^\tamusymbol$ Melanie Becker,\!$^\tamusymbol$ William D. Linch \textsc{iii},\!$^\tamusymbol$ \\[-4pt] \normalsize Stephen Randall,\!$^\berkeleysymbol$ and Daniel Robbins$^\albanysymbol$}
\date{}

\begin{document}
\maketitle

\thispagestyle{fancy}
\fancyfoot[C]{}
\fancyhead[R]{MI-TH-1620}

\begin{center}
\textit{$^\tamusymbol$George P. and Cynthia Woods Mitchell Institute for \\ Fundamental Physics and Astronomy, Texas A\&M University, \\[0.5cm] $^\berkeleysymbol$Department of Physics \\ University of California, Berkeley, \\[0.5cm] $^\albanysymbol$Department of Physics \\ University at Albany}
\end{center}
\vspace{5mm}

\abstract{
We give a geometric description of supersymmetric gravity/(non-)abelian $p$-form hierarchies in superspaces with 4D, $N=1$ super-Poincar\'e invariance.
These hierarchies give rise to Chern-Simons-like invariants, such as those of the 5D, $N=1$ graviphoton and the eleven-dimensional 3-form but also generalizations such as Green-Schwarz-like/$BF$-type couplings.
Previous constructions based on prepotential superfields are reinterpreted in terms of $p$-forms in superspace thereby elucidating the underlying geometry.
This vastly simplifies the calculations of superspace field-strengths, Bianchi identities, and Chern-Simons invariants.
Using this, we prove the validity of a recursive formula for the conditions defining these actions for any such tensor hierarchy. 
Solving it at quadratic and cubic orders, we recover the known results for the $BF$-type and cubic Chern-Simons actions. 
As an application, we compute the quartic invariant $\sim AdAdAdA+\dots$ relevant, for example, to seven-dimensional supergravity compactifications.
}

\newpage
\thispagestyle{fancy}
\fancyfoot[C]{}
\fancyhead[R]{}

\hypertarget{toc}{}
\noindent \hrulefill
\vspace*{-0.65cm}
\renewcommand*\contentsname{{\large Contents}}
\tableofcontents
\vspace*{-0.05cm}
\noindent \hrulefill

\pagestyle{fancy}
\fancyfoot[C]{--- ~~\hyperlink{toc}{\thepage}~~ ---}
\pagenumbering{arabic}

\section{Introduction}
\setcounter{equation}{0}
\label{sec:intro}

Gravitational tensor hierarchies are a common feature of gauged supergravity compactifications as they result from the reduction of $p$-forms in the component spectrum that are charged under the higher-dimensional superdiffeomorphisms \cite{deWit:2005hv,deWit:2008ta,deWit:2008gc, Bergshoeff:2009ph, Greitz:2013pua, Palmkvist:2013vya, Howe:2015hpa}.
Upon compactification, some of the components of the gravitino generally become massive but leave behind massless non-abelian gauge fields from mixed components of the frame and their superpartners.
What remains is a hierarchy of differential forms of various spacetime degrees, all charged under the residual diffeomorphisms compatible with the splitting of the compactified spacetime.
Further decoupling this structure from the lower-dimensional supergravity fields, one is left with a hierarchy of $p$-forms charged under the non-abelian gauge algebra of diffeomorphisms of the compactification manifold.

Such hierarchies of $p$-form fields, or ``tensor hierarchies'' as they have come to be known, come in various forms including abelian, non-abelian, and gravitational hierarchies.
The simplest such hierarchy arises in any theory containing a $p$-form field ($p>0$) on a product spacetime $X\times Y$:
Here the de Rham differential $d \to d_X + d_Y$ becomes a sum and the $p$-form splits into a collection of $p$-forms on $X$ valued in $q$-forms on $Y$. (In this case the ``hierarchy'' structure is encoded in the ordinary de Rham complex on $Y$.)

In particular, any dimensional reduction of a theory containing a $p$-form field will give rise to a tensor hierarchy of this type.
In the typical situation, the $p$-form fields in question are generalizations of the 1-form field of Mawell theory:
Abelian gauge transformations of such fields are exterior derivatives of $(p-1)$-form gauge parameters and the exterior derivative of such a potential is an invariant $(p+1)$-form field-strength.
The presence of such abelian $p$-forms is typical in extended supergravity and higher-dimensional supergravity theories:
These theories have additional spinor degrees of freedom arising from a larger gravitino, and the additional spin-0 and spin-1 degrees of freedom of the ``gravi-$p$-forms'' are needed to balance this without introducing more spin-2.
Well-known examples include the graviphoton ($+$ scalar) of 4D, $N=2$ supergravity and the 3-form of eleven-dimensional supergravity.

Gravitational tensor hierarchies arise naturally, then, in extended supergravity and higher-dimensional supergravity theories on product manifolds.
More generally, the background spacetime can have the structure of a non-trivial bundle over $Y$, in which case the mixed components of the graviton become the Kaluza-Klein gauge field for the algebra of diffeomorphisms on $Y$.
The reduced components of the gravi-$p$-forms are charged under this non-abelian gauge algebra: In addition to their usual abelian $p$-form transformation, they transform as matter fields under $Y$ diffeomorphisms.

The structure of the gravitational tensor hierarchy can be generalized by replacing the collection of dimensionally reduced forms with a more general set not necessarily resulting from any dimensional reduction. Maps between these new forms must be defined to replace the de Rham differential.
Provided this is done in a manner compatible with the de Rham complex of forms on $X$, there results an abelian tensor hierarchy of forms on $X$ with values in this new complex.

Abstracting further, the algebra of diffeomorphisms on $Y$ can then be replaced with a general non-abelian gauge algebra provided a representation is assigned to each degree in the new hierarchy. The action of this algebra should satisfy certain ``equivariance'' conditions with respect to the de Rham differential on $X$ and the maps of the $p$-form hierarchy.
These conditions can be interpreted as gauging the abelian hierarchy with respect to the new non-abelian gauge algebra;
such hierarchies are referred to as ``non-abelian tensor hierarchies''.\footnote{The conditions defining such a general non-abelian tensor hierarchy were formulated in \cite{Samtleben:2011fj} in an attempt to construct six-dimensional superconformally invariant gauge theories  (see also \cite{Samtleben:2012mi, Samtleben:2012fb}). The mathematical structure of these models was investigated further in \cite{Saemann:2012uq, Palmer:2013pka}.}
It is important to emphasize that the $p$-forms of the hierarchy transform linearly under the non-abelian group. Despite the terminology, the non-abelian aspect of the gauge structure is only that of the gauge field with the ``tensors'' transforming as matter fields.

Motivated by applications to supergravity compactifications, the defining conditions of such hierarchies were reformulated in \cite{Becker:2016rku} and interpreted as coming either from closure of the algebra of abelian and non-abelian gauge transformations or from the requirement that there exist enough gauge-covariant field-strengths.
Assuming both of these conditions, the resulting structure is that of a double chain complex that extends the superspace de Rham complex and is equivariant under the action of the non-abelian gauge group.
Being a differential chain complex in superspace, such theories naturally define supersymmetric characteristic classes, provided the appropriate traces are supplied.
In particular, it is possible to define the analogs of Maxwell/Yang-Mills invariants, higher Chern/Pontryagin classes, and Chern-Simons/$BF$-type invariants.

To better understand the four-dimensional phenomenology of such theories \cite{Bergshoeff:2009ph, deWit:2009zv, Hartong:2009az}, we first embedded the abelian \cite{Becker:2016xgv} and later the non-abelian \cite{Becker:2016rku} tensor hierarchies into flat, 4D, $N=1$ superspace and explicitly constructed their manifestly supersymmetric invariants.
The field-strengths of such hierarchies satisfy their Bianchi identities identically\footnote{They are given explicitly in terms of off-shell ``prepotential'' superfields. Such a (finite) set of prepotentials exists only because we have chosen to work in a superspace admitting no more than four real supercharges.}
and can be used to construct the usual Maxwell-type actions.
The superspace differential operators involved in the Bianchi identities turn out to be the adjoints of those appearing in the gauge transformations of the super-$p$-form potentials.
Because of this, the potentials and field-strengths can be used to construct a quadratic $BF$-type invariant.
By an abuse of language, we will refer to this as a quadratic Chern-Simons-type invariant for reasons that will hopefully become clear if they are not already.

In order to construct cubic and higher-order Chern-Simons-like actions (including the dimensional reductions of actual abelian Chern-Simons invariants), what is needed is a set of composite superfields constructed from the field-strengths that satisfy the same constraints ({\it i.e.}\ Bianchi identities) as the field-strengths themselves.
Such a construction of the Chern-Simons terms suffices since the inhomogeneous part of the $p$-form transformations is abelian.
Finding this set of composite superfields and checking the constraints requires considerable effort in the prepotential formulation since the constraints are not linear in superspace derivatives.
Indeed, it is not clear {\em a priori} that such a set of composites exists even in the abelian version of the hierarchy. 
In the non-abelian case the required interplay between hierarchy identities, gauge-covariant superspace $D$-algebra identities, and Bianchi identities seems miraculous.

This work originated in the desire to understand this ``miracle'' and to obviate the cumbersome calculus of the prepotential formalism by reinterpreting it in superspace differential-geometric terms. In such a formulation, the field-strengths are specific Lorentz-irreducible parts of super-$(p+1)$-forms \cite{Gates:1980ay}. The complicated Bianchi identities they satisfy are relations ``descendant'' from the condition that the superforms be closed (or exact by the Poincar\'e lemma).
Since the superspace de Rham operator is a graded derivation, the wedge product of closed forms is closed and the miraculous cancellations of the prepotential formalism would just be descendants of this trivial fact.
Finally, it was imagined that the Chern-Simons-like actions would simply be the integral of the (pullbacks of the) higher-dimensional super-Chern-Simons form.

As it turns out, this interpretation is overly-simplistic for two reasons. 
The first is that the closed superforms that define {\em irreducible} representations of the super-Poincar\'e algebra do not form a ring under multiplication:
To get an irreducible supermultiplet from a superform, many parts must be set to zero as conventional constraints (similarly to what is done to the torsion in superspace supergravity theories). 
On the other hand, when two lower-degree forms are wedged together they will generally give contributions violating these conditions and ruin the picture sketched above. 
This problem can be circumvented by defining an improved form in the same cohomology class that satisfies the original conditions on the irreducible superform.\footnote{By ``irreducible'' we will always mean as a representation of the super-Poincar\'e algebra. In particular, a form can be both composite ({\it i.e.}\ constructed by wedging non-composite forms) and irreducible ({\it i.e.}\ it satisfies the same constraints as the non-composite form of the same degree).}

The second complication is that the Chern-Simons action is defined by a form that is not closed whereas the ``ectoplasm'' method used to construct supersymmetric actions specifically requires the use of closed forms \cite{Gates:1997kr,Gates:1998hy}. 
Fortunately, it is known how to handle this situation \cite{Butter:2012ze, Kuzenko:2013rna}:
In addition to the Chern-Simons form $C$ constructed by wedging superform potentials and field-strengths, one constructs a second, inequivalent form $K$ that is both manifestly gauge invariant and satisfies $dK = dC$.
(That such a form exists is a phenomenon called ``Weil triviality'' \cite{Bonora:1986xd}, cf.\ \S\ref{S:WeilTriviality}.)
This gives a closed superform $J = C-K$ which, in turn,
defines the superspace completion of the component Chern-Simons action by the ectoplasm procedure (cf.\ \S\ref{S:Ectoplasm}).

Despite these complications to the na\"ive geometrization of the prepotential hierarchy, we will find that it is possible to give a recursive formula for the composite superfields defining the Chern-Simons invariants of any non-abelian tensor hierachy of the type defined in reference \cite{Becker:2016rku}.
Furthermore, it is possible to solve these recursion relations to obtain all of the superspace invariants explicitly.
We demonstrate this in detail by reproducing the cubic invariant found in \cite{Becker:2016rku} and deriving a new quartic invariant.
The former was recently used (in conjunction with a superspace Hitchin functional) to derive the scalar potential of eleven-dimensional supergravity in backgrounds admitting a (not necessarily closed) $G_2$ structure \cite{Becker:2016edk}.
The quartic invariant would be a main ingredient in a similar analysis for seven-dimensional supergravity backgrounds \cite{Pernici:1984xx}.

\paragraph{Outline}
We begin in section \ref{S:PrepotentialNATH} with a review of the non-abelian tensor hierarchy in the prepotential formulation and use this to describe the problem of constructing Chern-Simons-like gauge-invariant superspace actions.
This pre-geometrical description is reformulated in terms of super-$p$-forms in section \ref{S:pForms} where the composite superfields appearing in the construction of the (secondary) characteristic classes are interpreted as products of closed superforms.
In section \ref{S:Ectoplasm} we relate these composite superforms to supersymmetric invariants by way of ``ectoplasm".
This method takes as input a closed superform of spacetime degree four and returns a chiral superspace integral.
As mentioned above, the product of closed irreducible superforms is not a closed irreducible superform.
In section \ref{S:WeilTriviality}, we construct a gauge-invariant composite superform with which we modify the original composite superform to obtain a closed, irreducible, composite superform.
Applying the ectoplasm method, this corrected composite form gives the Chern-Simons-like superspace action.
Since the construction is in terms of superforms, the result is manifestly supersymmetric and gauge-invariant by the same logic as that for bosonic Chern-Simons invariants.
In section \ref{S:AllCS}, we use this technology to derive and solve a recursion formula for all Chern-Simons-like actions for any non-abelian tensor hierarchy.
We summarize our conclusions in section \ref{S:Conclusions}.

As we will see throughout our presentation, the use of superforms streamlines the construction of geometric invariants and simplifies or obviates many cumbersome and delicate calculations. 
We have attempted to make this paper self-contained but have been necessarily brief in our review of superform methods. 
A pedagogical introduction to superforms in the context of tensor hierarchies may be found in reference \cite{Randall:2016zpo}.

\section{Prepotential Formalism}
\label{S:PrepotentialNATH}

The papers \cite{Becker:2016xgv,Becker:2016rku} were motivated by the goal of writing a supersymmetric theory in $D>4$ dimensions, particularly eleven-dimensional supergravity, in 4D $N=1$ language. The results obtained were more general, with no assumptions being made about whether the four-dimensional tensor hierarchy had been obtained from a higher-dimensional theory or not. In the present work we will not be as careful to maintain this full generality, although this choice is primarily made to keep the notation simple. Instead we will implicitly assume that the four-dimensional tensor hierarchy arises from a $p$-form in $D$ dimensions, where the $D$-dimensional theory is being put on a background $\R^4\times M$, with $M$ a $(D-4)$-dimensional internal space. Note that $p$ should be odd in order for us to have a non-trivial Chern-Simons action.

In this case, the bosonic four-dimensional tensor hierarchy is comprised of axions $a$, which are zero-forms in spacetime and $p$-forms on $M$, spacetime one-forms $A_a$ which are $(p-1)$-forms on $M$, spacetime two-forms $B_{ab}$ valued in internal $(p-2)$-forms, spacetime three-forms $C_{abc}$ valued in internal $(p-3)$-forms, and spacetime four-forms $D_{abcd}$ valued in internal $(p-4)$-forms.  Note that if $p=3$ there simply are no $D_{abcd}$ fields, and if $p=1$ there are only axions and 1-forms. These forms can be multiplied, using the wedge product for forms on $M$, and if $D=n(p+1)-1$ for some $n>1$, then we can construct a $D$-dimensional Chern-Simons action by wedging one potential and $n-1$ field-strengths and integrating the resulting $D$-form over $\R^4\times M$. By integrating just over $M$, we get a 4D Chern-Simons action for the tensor hierarchy.

Additionally, if we are reducing a supergravity theory in $D$ dimensions, then we can also incorporate the 4D gauge fields coming from off-diagonal components of the $D$-dimensional metric. These are spacetime one-forms which are tangent vectors on $M$, and their corresponding non-abelian gauge group is the group of diffeomorphisms on $M$ (whose Lie algebra can be identified with $\G(TM)$, the space of vector fields on $M$ with the usual Lie bracket).

In \cite{Becker:2016rku}, it was explained how to embed these structures into 4D, $N=1$ superfields. The non-abelian gauge vectors $\mcA_a$ were promoted to $TM$-valued super-1-forms $\mcA_A$ (with the lowest components of the superfield $\mcA_a$ matching the bosonic fields of the same name) which were used to build gauge covariant super-derivatives $\D_A=\{\D_a,\D_\al,\bar{\D}_{\dot\al}\}$ by
\be
\label{E:Superconnection}
\D_a=\p_a-(\mcL_\mcA)_a,\quad\D_\al=D_\al-(\mcL_\mcA)_\al,\quad\bar{\D}_{\dot\al}=\bar{D}_{\dot\al}-(\mcL_\mcA)_{\dot\al}.
\ee
Here $\mcL_\mcA$ is the Lie derivative along the vector field $\mcA$ which acts on differential forms of the internal space.

The bosonic fields of the hierarchy are embedded in superfield prepotentials as \cite{Gates:1980ay}
\begin{subequations}
\label{subeqns:FormEmbed}
\begin{align}
a &= \hlf(\Phi+\bar{\Phi})\big|,\\
\label{subeqn:AEmbed}
A_a &= -\frac{1}{4}(\bar{\s}_a)^{\dot\al\al}[\D_\al,\bar{\D}_{\dot\al}] V\big|,\\
B_{ab} &= -\frac{i}{2}[(\s_{ab})_\al^{\hph{\al}\beta}\D^\al\Sigma_\beta-(\bar{\s}_{ab})^{\dot\al}_{\hph{\dot\al}\dot\beta}\bar{\D}_{\dot\al}\bar{\Sigma}^{\dot\beta}]\big|,\\
C_{abc} &= \frac{1}{8}\e_{abcd}(\bar{\s}^d)^{\dot\al\al}[\D_\al,\bar{\D}_{\dot\al}] X\big|,\\
D_{abcd} &= \frac{i}{8} \epsilon_{abcd} (\D^2\G-\bar{\D}^2\bar{\G})\big|.
\end{align}
\end{subequations}
Here $\Phi$ is a covariantly chiral superfield valued in $p$-forms on $M$, $V$ is a real superfield valued in $(p-1)$-forms, $\Sigma_\al$ is a covariantly chiral spinor superfield valued in $(p-2)$-forms, $X$ is a real superfield valued in $(p-3)$-forms, and $\G$ is a covariantly chiral superfield valued in $(p-4)$-forms.  A vertical slash means that we take the lowest component of the superfield.

The non-abelian gauge transformations take the form $\d S=\mcL_\la S$ for any of the hierarchy superfields $S$, where $\la$ is a $TM$-valued real superfield parameterizing the non-abelian gauge transformations and the super-1-form $\mcA_A$ transforms as $\d\mcA_A=\D_A\la$ (with the Lie derivative inside of $\D_A$ now acting on vector fields). The gauge-invariant part of the non-abelian vectors is captured by a $TM$-valued covariantly chiral spinor superfield $\mcW_\al$ satisfying $\D^\alpha \mcW_\alpha = \bar{\D}_\ad \bar{\mcW}^\ad$.

The bosonic gauge transformations from the hierarchy now lift to superfield gauge transformations parameterized by chiral $\La$, real $L$, chiral spinor $\Upsilon_\al$, real $\Xi$, and chiral $\Pi$ superfields valued in $(p-1)$-, $(p-2)$-, $(p-3)$-, $(p-4)$\mbox{-,} and $(p-5)$-forms respectively.  Note that if $p=3$, as in the reduction from eleven-dimensional supergravity, then the last two gauge parameters do not appear.  The transformations are
\begin{subequations}
\label{eqs:Transfos}
\begin{align}
\d\Phi &= \p\La,\\
\d V &= \frac{1}{2i}(\La-\bar{\La})-\p L,\\
\d\Sigma_\al &= -\frac{1}{4}\bar{\D}^2\D_\al L+\p\Upsilon_\al+(\iota_\mcW)_\al\La,\\
\d X &= \frac{1}{2i}(\D^\al\Upsilon_\al-\bar{\D}_{\dot\al}\bar{\Upsilon}^{\dot\al})-\p\Xi-\Om(\iota_\mcW,L),\\
\d\G &= -\frac{1}{4}\bar{\D}^2\Xi+\p\Pi+(\iota_\mcW)^\al\Upsilon_\al.
\end{align}
\end{subequations}
Here $\p$ is the exterior derivative acting on differential forms on $M$, and $\iota_v$ is contraction of a form by a vector field $v$. The object $\Om(\cdot,\cdot)$ is the so-called Chern-Simons superfield,
\be
\Om(\psi,S) \ceq \psi^\al\D_\al S+\bar{\psi}_{\dot\al}\bar{\D}^{\dot\al}S+\hlf(\D^\al\psi_\al+\bar{\D}_{\dot\al}\bar{\psi}^{\dot\al}) S,
\ee
which takes as input a covariantly chiral spinor superfield $\psi^\al$ and a real superfield $S$.\footnote{Its name derives from the property (in flat superspace for simplicity) 
$\bar D^2\Om(\psi,S)=\psi^\al\bar D^2D_\al S+\hlf( D^\al\psi_\al-\bar D_{\dot\al}\bar{\psi}^{\dot\al}) S$ 
so if $\psi_\alpha \to W_\alpha = -\tfrac14 \bar D^2 D_\alpha V$,
then $-\tfrac14\bar D^2 \Omega(W, V) = W^\alpha W_\alpha$ gives the superspace version of $d \Omega = F \wedge F$. In terms of superforms, this corresponds to deforming the 3-form field strength $H \to  dB +X$ \cite{Siegel:1995px} by the Chern-Simons super-3-form $X=\mathrm{tr}\left(\tfrac12 AdA+\tfrac13A^3\right)$ \cite{Nishino:1991sr, Gates:1991qn}. (Applications to the chiral anomaly in superspace were studied in \cite{Gates:2000dq, Gates:2000gu}.)
}

It is possible to construct a set of gauge-invariant (under the hierarchy transformations (\ref{eqs:Transfos}), and covariant under the non-abelian gauge transformations) field-strength superfields,
\begin{subequations}
\label{eqs:SuperfieldStrengths}
\begin{align}
E &= \p\Phi,\\
U &= \frac{1}{2i}(\Phi-\bar{\Phi})-\p V,\\
W_\al &= -\frac{1}{4}\bar{\D}^2\D_\al V+\p\Sigma_\al+(\iota_\mcW)_\al\Phi,\\
H &= \frac{1}{2i}(\D^\al\Sigma_\al-\bar{\D}_{\dot\al}\bar{\Sigma}^{\dot\al})-\p X-\Om(\iota_\mcW,V),\\
G &= -\frac{1}{4}\bar{\D}^2X+\p\G+(\iota_\mcW)^\al\Sigma_\al.
\end{align}
\end{subequations}
Of these, $E$, $W_\al$, and $G$ are chiral superfields, whereas $U$ and $H$ are real superfields.
They satisfy the Bianchi identities
\begin{subequations}
\label{eqs:Bianchis}
\begin{align}
0 &= \p E,\\
0 &= \frac{1}{2i}( E-\bar{E})-\p U,\\
0 &= -\frac{1}{4}\bar{\D}^2\D_\al U+\p W_\al+(\iota_\mcW)_\al E,\\
0 &= \frac{1}{2i}(\D^\al W_\al-\bar{\D}_{\dot\al}\bar{W}^{\dot\al})-\p H-\Om(\iota_\mcW,U),\\
0 &= -\frac{1}{4}\bar{\D}^2H+\p G+(\iota_\mcW)^\al W_\al.
\end{align}
\end{subequations}
In the next section we will relate the repeating patterns in (\ref{eqs:Transfos}), (\ref{eqs:SuperfieldStrengths}), and (\ref{eqs:Bianchis}) to the action of the superspace de Rham operator on superforms.

We can now write a candidate super-Chern-Simons action as
\be
\label{eq:CompositeAction}
S_{SCS}=\Real \left[ i\int d^4xd^2\t\Tr(\Phi g+\Sigma^\al w_\al+\G e) \right] +\int d^4xd^4\t\Tr( Vh-Xu),
\ee
where $e$, $u$, $w_\al$, $h$, and $g$ are composite superfields built out of the field-strengths (\ref{eqs:SuperfieldStrengths}), with $e$, $w_\al$, and $g$ chiral, and $u$ and $h$ real.  Here the $\Tr$ represents an integration of the internal $(D-4)$-form over the internal space (and for cases other than those coming from dimensional reduction, it is possible to assign a suitably generalized meaning).
Then the conditions for gauge invariance of $S_{SCS}$ under the hierarchy transformations (\ref{eqs:Transfos}) are the ``descent relations''
\begin{subequations}
\label{eqs:DescentRelations}
\begin{align}
0 &= \p e,\\
0 &= \frac{1}{2i}( e-\bar{e})-\p u,\\
0 &= -\frac{1}{4}\bar{\D}^2\D_\al u+\p w_\al+(\iota_\mcW)_\al e,\\
0 &= \frac{1}{2i}(\D^\al w_\al-\bar{\D}_{\dot\al}\bar{w}^{\dot\al})-\p h-\Om(\iota_\mcW,u),\\
0 &= -\frac{1}{4}\bar{\D}^2h+\p g+(\iota_\mcW)^\al w_\al.
\end{align}
\end{subequations}
In other words, the composite fields must satisfy the same Bianchi identities (\ref{eqs:Bianchis}) as the field-strengths themselves. This means that to build a quadratic super-Chern-Simons action, we should simply take $e=E$, $u=U$, $w_\al=W_\al$, $h=H$, and $g=G$. To build actions that are higher order in the number of fields apparently requires significantly more work, and in \cite{Becker:2016xgv,Becker:2016rku} this was done to cubic order essentially by writing down all possible terms which could appear in the composites and then fixing the relative coefficients by solving (\ref{eqs:DescentRelations}).

This concludes our review of our previous results on non-abelian tensor hierarchies and their Chern-Simons invariants in 4D, $N=1$ superspace.
The main result of this paper can now be stated precisely as the explicit construction of any Chern-Simons action of the form (\ref{eq:CompositeAction}).
The lemma we will need to establish is the following
\begin{claim}
\label{Th:Recursive}
Suppose that we have constructed the composites $e_n$, $u_n$, $w_n^\al$, $h_n$, and $g_n$ that solve the descent equations (\ref{eqs:DescentRelations}) and are of order $n$ in the field-strength superfields.
Then the composite superfields
\begin{subequations}
\label{eqs:Recursion}
\begin{align}
e_{n+1} &= Ee_n,\\
u_{n+1} &= \hlf( E+\bar{E}) u_n+\hlf U( e_n+\bar{e}_n),\\
w_{n+1}^\al &= Ew_n^\al+W^\al e_n+\frac{i}{4}\bar{\D}^2(\D^\al Uu_n-U\D^\al u_n),\\
h_{n+1} &= \hlf( E+\bar{E}) h_n+\hlf H( e_n+\bar{e}_n)+\Om(w_n,U)+\Om(W,u_n)\\
& \quad -i\D^\al U(\iota_\mcW)_\al u_n-i(\iota_\mcW)^\al U\D_\al u_n +i\bar{\D}_{\dot\al}U(\iota_{\bar{\mcW}})^{\dot\al}u_n+i(\iota_{\bar{\mcW}})_{\dot\al}U\bar{\D}^{\dot\al}u_n,\cr
g_{n+1} &= Eg_n+Ge_n+W^\al w_{n\,\al}+\frac{i}{4}\bar{\D}^2( Hu_n-Uh_n),
\end{align}
\end{subequations}
of order $n+1$ also satisfy the descent relations.
\end{claim}

The most straightforward proof of Claim \ref{Th:Recursive} is to substitute the expressions (\ref{eqs:Recursion}) into the descent relations (\ref{eqs:DescentRelations}) and verify that all terms cancel using the Bianchi identities (\ref{eqs:Bianchis}).
This, however, does not elucidate the structure of the recursion relation nor the underlying reason the descent equations admit a non-trivial solution in the first place.
Instead, a constructive proof of this claim will be given in section \ref{S:AllCS} once we have developed the necessary supergeometry.

\section{Superforms}
\label{S:pForms}

In this section we geometrize the prepotential formalism of the previous section by recasting it in terms of differential forms in superspace \cite{Gates:1980ay}.
We define a superform $\omega$ of degree $p$ by the na\"ive extension of the coordinate expression of a bosonic $p$-form,
\begin{align}
\label{E:pForm1}
\omega = \frac1{p!} dz^{M_p} \wedge \dots \wedge dz^{M_1} \omega_{M_1\dots M_p}(z).
\end{align}
Here $z^M=(x^m, \theta^\mu, \bar \theta^{\dot \mu})$ stands for the Cartesian super-coordinates
and $dz^M \wedge dz^N = - (-1)^{MN} dz^N \wedge dz^M$ is the graded wedge product.
The de Rham operator
\begin{align}
\label{E:deRham}
d = dz^M {\partial\over \partial z^M} = dz^M \partial_M
\end{align}
maps super-$p$-forms to super-$(p+1)$-forms with
\begin{align}
\label{E:dpForm}
d\omega &= \frac1{p!} dz^{M_p} \wedge \dots \wedge dz^{M_1} \wedge dz^N \partial_{N} \omega_{M_1\dots M_p}(z) \cr
	& = \frac1{p!} dz^{M_p} \wedge \dots \wedge dz^{M_1} \wedge dz^N \partial_{[N} \omega_{M_1\dots M_p]}(z).
\end{align}
Here ${}_{[\ldots]}$ denotes graded anti-symmetrization of indices. The partial derivatives super-commute $[\partial_M, \partial_N] = \partial_M \partial_N  - (-1)^{MN} \partial_N \partial_M = 0$.
Thus, the super-de Rham operator is a differential and we can construct the superspace analog of the de Rham complex
\begin{align}
\label{E:deRhamComplex}
\Omega^\bullet  ~:~ 0 \longrightarrow \Omega^0
	\stackrel {d} \longrightarrow
\Omega^1
	\stackrel {d} \longrightarrow
\Omega^2
	\stackrel {d} \longrightarrow
\Omega^3
\longrightarrow \dots
\end{align}

Gauge potentials $A$ are defined by closed forms $dF = 0 \,\Rightarrow\, F = dA$ through the Poincar\'e lemma
and we would like to extend this to superspace gauge potentials.
The na\"ive solution $F_{M_1\dots M_p} = p!\, \partial_{[M_1}A_{M_2\dots M_p]}$ does not define a linear representation of the supersymmetry algebra because the fermionic coordinate derivatives do not commute with the supercharges.
The solution to this problem is to pass to a super-covariant basis of forms by introducing flat superspace vielbeins $E_A{}^M$ and their inverses $E_M{}^A$
\begin{align}
\label{E:deRham2}
d = dz^M E_M{}^A E_A{}^M\partial_M  = E^A D_A.
\end{align}
Similarly, we rearrange
\begin{align}
\label{E:pForm2}
\omega = \frac1{p!} E^{A_p} \wedge \dots \wedge E^{A_1} \omega_{A_1\dots A_p}(z).
\end{align}

The flat superspace covariant derivatives $D_A = (D_\alpha, \bar D_{\dot \alpha}, \partial_a)$ commute with the supercharges but now the frames carry torsion $T^A = dE^A$. This changes the formula (\ref{E:pForm1}) for the exterior derivative to the covariant version
\begin{align}
\label{E:dpForm2}
d\omega &= \frac1{p!} E^{A_p} \wedge \dots \wedge E^{A_1} \wedge E^B
	\left(
		D_{[B} \omega_{A_1\dots A_p]}(z)
		+ \frac12 T_{[BA_1}{}^C \omega_{|C|A_2\dots A_p]}(z)
	\right)\!.
\end{align}
Here ${}_{|\dots|}$ indicates that ${}_{\dots}$ is to be omitted from the anti-symmetization.

The collection of superfields $\omega_{A_1\dots A_p}(z)$ is taken to be graded-anti-symmetric so that
the components of a $p$-form $\om$ which are dimension-$(\tfrac12(t+u)+v)$ are superfields of the form
\be
\om_{\al_1\cdots\al_t\dot\al_1\cdots\dot\al_ua_1\cdots a_v},\qquad t+u+v=p.
\ee
They are symmetric under interchange of any two spinor indices, but anti-symmetric under any other exchange of indices.  To automatically keep track of these symmetry properties, it can be useful to introduce commuting spinor variables $s^\al$ and $\bar{s}^{\dot\al}$, and anti-commuting vector variables $\psi^a$, which allows us to use more compact notation,
\be
\label{E:Janktification}
\om_{s\cdots s\bar{s}\cdots\bar{s}\psi\cdots\psi}=\om_{\al_1\cdots\al_t\dot\al_1\cdots\dot\al_ua_1\cdots a_v}s^{\al_1}\cdots s^{\al_t}\bar{s}^{\dot\al_1}\cdots\bar{s}^{\dot\al_u}\psi^{a_1}\cdots\psi^{a_v}.
\ee
(Note that since the spinor indices are symmetrized, we can have $p$-forms with $p>4$ in four dimensions. Such ``over-the-top forms'' appear in closely-related hierarchies \cite{Greitz:2011vh, Greitz:2012vp, Howe:2015hpa}.)

In flat 4D, $N=1$ superspace there is only one non-vanishing torsion $T_{\alpha \dot \alpha}{}^a = -2i (\sigma^a)_{\alpha \dot \alpha}$.
Thus the independent components appearing in the exterior derivative of a $p$-form (\ref{E:dpForm2}) can we written as
\begin{align}
\label{E:Jank}
( d\om)_{s\cdots s\bar{s}\cdots\bar{s}\psi\cdots\psi} &=
	( -1)^{t+u}v\p_\psi\om_{s\cdots s\bar{s}\cdots\bar{s}\psi\cdots\psi}
	+tD_s\om_{s\cdots s\bar{s}\cdots\bar{s}\psi\cdots\psi}+u\bar{D}_{\bar{s}}\om_{s\cdots s\bar{s}\cdots\bar{s}\psi\cdots\psi} \non\\
& \qquad +2i( -1)^{p+1}tu(\s^a)_{s\bar{s}}\om_{s\cdots s\bar{s}\cdots\bar{s}\psi\cdots\psi a},
\end{align}
where $t+u+v=p+1$, is the degree of $d\om$.
As in the ordinary case, we say that $\om$ is closed if $d\om=0$, and $\om$ is exact if $\om=d\eta$.

We have ordered the terms so that the superfield in the first has the lowest dimension, the next two have dimension one-half higher, and the last has dimension one higher.
Solving this covariant closure condition $dF=0$ now gives covariant components for the potential $A$ but there are {\em way} too many components in a general potential $p$-form to define an irreducible representation of the super-Poincar\'e algebra. (An unconstrained superfield is reducible and we have a large collection of such superfields.)
This is solved by setting the lower-dimensional components of the covariant field strength $F$ to zero.
This is analogous to the torsion constraints in superspace supergravity: When this doesn't trivialize the form, the conditions give a covariant superfield because the constraints are covariant.
The first component that is not set to vanish ({\it i.e.}\ that with the lowest dimension) must then satisfy the relation
\begin{align}
\label{eq:cocycle}
(\s^a)_{s\bar{s}}\om_{s\cdots s\bar{s}\cdots\bar{s} a\psi\cdots\psi} = 0,
\end{align}
which follows from (\ref{E:Jank}) with the lower-dimension components set to zero.\footnote{The exception is when the dimension-$\tfrac{p}{2}$ components are non-zero, which we have for the 1-form field-strengths (cf. table \ref{T:superforms}).}
In table \ref{T:superforms} we give the solutions to this condition for all 4D, $N=1$ $p$-form field-strengths \cite{Gates:1980ay}.

One then inserts this component into the next-higher dimension closure condition and solves the next-higher component in terms of $D$ and $\bar D$ on the first and so on. This gives conditions on the superfields in the table below.\footnote{This seemingly {\it ad hoc} (and potentially inconsistent, if there is no solution to \eqref{eq:cocycle}) procedure can be justified by interpreting the de Rham operator in terms of the Chevalley-Eilenberg differential of the super-translation algebra. In this context, the procedure is computing the Lie algebra cohomology of this superalgebra with values in the module of superfields \cite{Linch:2014iza}.
} 

\begin{table}[h]
{\renewcommand{\arraystretch}{1.3} 
\begin{center}
\begin{tabular}{|c|c|c|c|c|c|}
\hline
$p$ & strength & constraints & prepotential & top component  \cr
\hline
0 & $F_\alpha = D_\alpha U$
	& $\bar D^2 D U=0$
	& $U = \Phi + \bar \Phi$
	& $F_a = D\sigma_a \bar D U +\mathrm{c.c.}$ \cr
1 & $F_{\alpha a} = (\sigma_a \bar W)_\alpha$
	& $D \bar W=0$ \& $\bar D \bar W=DW$
	& $W = \bar D^2 D V$
	& $F_{ab} = D\sigma_{ab} W +\mathrm{c.c.}$ \cr
2 & $F_{\alpha\dot\alpha a} = (\sigma_a)_{\alpha \dot \alpha} H $
	& $H=\bar H$ \& $D^2 H =0$
	& $H = D\Sigma + \bar D \bar\Sigma$
	& $F_{abc} = \epsilon_{abcd} D\sigma^d\bar D H +\mathrm{c.c.}$ \cr
3 & $F_{\alpha\beta ab} = (\sigma_{ab})_{\alpha \beta} \bar G $
	& $\bar D G=0$
	& $G = \bar D^2 X$
	& $F_{abcd} = \epsilon_{abcd} D^2G +\mathrm{c.c.} $ \cr
\hline
\end{tabular}
\end{center}
} 
\begin{caption}{Embedding of $p$-forms in closed superforms}
\label{T:superforms}
\footnotesize
The component $p$-forms are embedded into closed super-$(p+1)$-form field-strengths $F$ as originally shown in \cite{Gates:1980ay}. Each field-strength can be written in terms of an invariant scalar or spinor superfield. These satisfy constraints that can be solved in terms of prepotentials. Numerical coefficients are neglected in this table for simplicity but can be found in section \ref{S:PrepotentialNATH} (for the constraints and prepotentials) and in appendix \ref{S:MoreForms} (for the superform components).
\end{caption}
\end{table}

Based on table \ref{T:superforms}, we again see a pattern (as was mentioned in the previous section) between the structures of the constraints and prepotential solutions. Although not included in the table, this pattern extends to the gauge variations as well. It is not always appreciated that these rhyming structures are simply consequences of nilpotency at various levels of the complex. In the original superforms paper \cite{Gates:1980ay} this is observed as prepotential solutions having a ``memory" of the gauge transformations and in \cite{Becker:2016xgv} it is remarked that there is a ``beautiful symmetry" between the constraints, solutions, and variations. This is not a coincidence and merely follows from $\omega = d \chi$ solving $d \omega = 0$ identically and $\delta \chi = d \sigma$ leaving $\omega$ invariant. However, when this is checked by crunching through $D$-calculus it appears quite a bit more impressive. This is because the linear closure condition $d \omega = 0$ is not necessarily linear in $D$'s when expressed in terms of superfields. Closure of the 3-form involves the quadratic operator $D^2$, while closure of the 1-form is cubic in $D$. This means that at each degree, the nilpotency of $d$ is in terms of non-trivial higher-order $D$-identities. For example, the prepotential solution for $W_\alpha$ in the super-de Rham complex works because
\begin{equation}
\label{eq:nilid}
	D^\alpha \bar{D}^2 D_\alpha = \bar{D}_\ad D^2 \bar{D}^\ad.
\end{equation}
Since this is equivalent to the nilpotency relation $d(dF) \equiv 0$ (for $\deg(F) = 1$), we did not need to know \eqref{eq:nilid} to write down the prepotential solution. In more complicated superspaces (\textit{e.g.}, higher-dimensional, $N > 1$, curved, \textit{etc.}) these identities are often significantly more complicated and ways to avoid having to rely on them are subsequently more valuable. Without this geometric perspective, checking things as simple as gauge covariance can become forbiddingly involved.

Returning to the matter at hand, the procedure above gives irreducible, off-shell representations of the Poincar\'e group.
Achieving these properties required that certain lower-dimensional components of the superform vanish.
This condition is not preserved by the wedge product, as is easy to see by considering the product of two 1-form field-strengths $F$ and $F'$:
The lowest non-vanishing components are $\mathbf F_{ss} = F_s F'_s$, $\mathbf F_{s\bar s} = F_s F'_{\bar s}$ and their conjugates.
This is in contradiction with the conditions for an irreducible 2-form field-strength since, as we see from the second row of table \ref{T:superforms}, the lowest-dimension non-vanishing component of an irreducible, closed 2-form is $F_{s \psi}$.

Contrary to the case of ordinary de Rham forms then, irreducible superforms do not give rise to a differential graded superalgebra.
In section \ref{S:Ectoplasm} we will construct superspace actions from closed irreducible 4-forms.
To apply this to composite forms we will have to address this apparent obstruction to irreducibility.

\subsection{Differential Supergeometry of Tensor Hierarchies}
\label{S:GeomNATH}

We are now in a position to complete the geometrization of the non-abelian tensor hierarchies reviewed in section \ref{S:GeomNATH}.
At the most abstract level such hierarchies are double complices of superspace de Rham forms (\ref{E:deRhamComplex}) with values in a differential complex of representations $GL(K^i)$ of some Lie algebra \cite{Becker:2016xgv}:
\begin{align}
\label{E:RepComplex}
K^\bullet  ~:~ 0 \longrightarrow K^0
	\stackrel {q} \longrightarrow
K^1
	\stackrel {q} \longrightarrow
K^2
	\stackrel {q} \longrightarrow
K^3
\longrightarrow \dots
\end{align}
When the Lie algebra is gauged, the de Rham operator $d \to \mathcal D$ acquires a connection and the new differential $q$ must commute (in the appropriately graded sense) with this covariant exterior derivative \cite{Becker:2016rku}.

A large family of realizations of this setup arise in compactification scenarios in which a higher-dimensional theory of differential forms is reduced on a super-vector bundle over a smooth bosonic base $Y$ of some dimension $n$.
Then the representation spaces $K^i = \Omega^{i}(Y)$ are the spaces of forms on $Y$ and the differential $q$ is the de Rham operator on $Y$.
The gauging is by the diffeomorphisms on $Y$ with the gauge field identified with the mixed components of the frame (with one leg in the tangent directions of the base and one in the superspace fiber).
In the apparently more general situation of the abstract hierarchy, the notation needed to keep track of the many ingredients defining the representation complex and gauging can be quite cumbersome.
To avoid this, we will proceed using the notation and language arising from compactifications.\footnote{This is much less of a restriction than it may initially seem since it applies to any situation in which the complex (\ref{E:RepComplex}) admits a free resolution \cite{cartan1956homological}.
In this case there will be an analog of $Y$ with its local coordinate derivatives and 1-forms so that we can continue to use the concepts and notation from compactifications for this new formal bosonic space.
}

A $q$-form of the abelian hierarchy is a sum of super-$p$-forms with values in $\Omega^{q-p}(Y)$ for $p=0,\dots, q$,
\begin{align}
\label{E:AbelianForm}
\Omega^q = \bigoplus_{p=0}^q \Omega^p(\mathbf R^{4|4}) \otimes \Omega^{q-p}(Y).
\end{align}
Denoting, as in section \ref{S:PrepotentialNATH}, the de Rham operator on $Y$ by $\partial$, the differential $Q$ on this abelian hierarchy is
\begin{align}
\label{E:AbelianDiff}
Q=d + q,
~~~\mathrm{with}~~~
q = (-1)^{p+1}\partial
~~~\mathrm{on}~~~
\Omega^p(\mathbf R^{4|4}) \otimes \Omega^{q-p}(Y).
\end{align}
Here $d$ is the superspace de Rham operator acting on superforms as defined by (\ref{E:Jank}) and
the alternating sign is needed to have $q$ anti-commute with $d$.
Redoing the analysis of closed $(q+1)$-form field-strengths for this complex is unnecessary since this is just the usual double complex construction in which $QF = 0$ is solved by $F=QA$ for some $q$-form $A$.\footnote{Explicitly, $F = \sum_{p=0}^{q+1} F_{[p,q+1-p]}$ with $F_{[p,q]}= dA_{[p-1,q]}+(-1)^{p+1} \partial A_{[p,q-1]}$ for $A= \sum_{p=0}^{q} A_{[p,q-p]}$.}
In terms of prepotentials, one sees that this is (\ref{eqs:SuperfieldStrengths}) with $\mcW^\al$ turned off.
We note that $F$ now also has a degree-zero part
\be
F^{(0)}=( QC)^{(0)}=\p C^{(0)}=\hlf( E+\bar{E}),
\quad \mathrm{where} \quad
E=\p\Phi.
\ee

Gauge symmetry again takes the form $\d C=Q\La$, where $\La^{(p-1)}$ is obtained from the expressions for $C^{(p-1)}$ by substituting the gauge parameter superfields $\La$, $L$, $\Upsilon^\al$, $\Xi$, and $\Pi$ in for the prepotentials $\Phi$, $V$, $\Sigma^\al$, $X$, and $\G$ respectively.  The invariance of the superforms $F^{(p+1)}$ again follows from nilpotence of the differential, $Q^2=0$.

Gauging the hierarchy means that we replace the superspace de Rham differential $d \to d_{\mathcal A}$ with the gauge-covariant exterior derivative defined by (\ref{E:Superconnection}). This is no longer a differential, since
\begin{align}
d_{\mathcal A}^2 = - \mathcal{L}_{\mathcal F},
\end{align}
where $\mathcal F$ is the non-abelian super-2-form field-strength
related to the superfield $\mathcal W$ of section \ref{S:PrepotentialNATH} by $\mathcal F_{s \psi} = s \sigma_\psi \bar {\mathcal W}$ (cf.\ table \ref{T:superforms}).
Using the compactification language, we may think of $\mathcal F$ as a super-2-form valued in vector fields on $Y$ so that it is sensible to contract it with forms on $Y$.
A differential can then be constructed as
\begin{align}
\label{E:NonAbelianDiff}
\mathcal Q=d_{\mathcal A} + q + \iota_{\mathcal F},
\end{align}
where the contraction operator is such that \cite{Becker:2016rku}
\begin{align}
q  \iota_{\mathcal F} +  \iota_{\mathcal F} q  = \mathcal L_{\mathcal F}.
\end{align}
Explicitly, the contraction term acts on superforms by
\begin{align}
\iota_{\mathcal F} \omega_{s\dots s\bar s \dots \bar s \psi\dots \psi}
	& =(-1)^{v+1}tv (\iota_{\mathcal F_{s\psi}}) \omega_{s\dots s\bar s \dots \bar s \psi\dots \psi}
	+(-1)^{v+1} uv (\iota_{\mathcal F_{\bar s\psi}}) \omega_{s\dots s\bar s \dots \bar s \psi\dots \psi} \notag\\
	& \quad +\tfrac12 (-1)^{p+1}v(v-1) (\iota_{\mathcal F_{\psi\psi}}) \omega_{s\dots s\bar s \dots \bar s \psi\dots \psi}.
\end{align}

Having constructed a covariant superspace differential (\ref{E:NonAbelianDiff}), the rest is straightforward. Gauge transformations, gauge-covariant field-strengths, and Bianchi identities are all given in terms of $\mathcal Q$.
We collect the explicit expressions for the superform gauge fields and field-strengths in appendix \ref{S:MoreForms}. The covariant constraints coming from superform closure are displayed in \eqref{eqs:Bianchis}. (To see how exactly these constraints arise, see \S5 of \cite{Randall:2016zpo}.)

\section{Ectoplasm}
\label{S:Ectoplasm}

The ectoplasm formalism \cite{Gates:1997kr,Gates:1998hy} is a way of constructing supersymmetric $D$-dimensional actions from closed, irreducible superforms of degree $D$. Specializing to $D=4$, let us first consider the case of standard super-de Rham cohomology. Suppose $J$ is a closed super-4-form and define
\be
\label{E:EctoAction}
S_J \ceq \tfrac{1}{24}\int d^4x\e^{abcd}J_{abcd}\big|.
\ee
This is supersymmetric because under a supersymmetry transformation we have
\begin{align}
\delta S_J & = \tfrac{1}{24} \varepsilon^{\hat{\al}}\int d^4x\e^{abcd} {Q}_{\hat{\al}}J_{abcd}\big|=\tfrac{1}{24} {\varepsilon}^{\hat{\al}}\int d^4x\e^{abcd}{D}_{\hat{\al}}J_{abcd}\big| \notag\\
&= \tfrac{1}{6} {\varepsilon}^{\hat\al}\int d^4x\e^{abcd}\p_aJ_{\hat{\al} bcd}\big|=0,
\end{align}
where for brevity we combined $\varepsilon^{\hat \alpha} Q_{\hat \alpha} \equiv \varepsilon^{\alpha} Q_{\alpha} + \bar \varepsilon_{\dot \alpha} Q^{\dot \alpha}$.
Here the second equality holds because $\bar{Q}$ and $\bar{D}$ differ by a spacetime derivative,
while the third equality holds by closure: $0=( dJ)_{\dot\al abcd}=\bar{D}_{\dot\al}J_{abcd}-4\p_{[a}J_{|\dot\al|bcd]}$.
Furthermore, if the lowest-dimension non-vanishing components of $J$
are $J_{\dot\al\dot\beta ab}=-4(\bar{\s}_{ab})_{\dot\al\dot\beta}J_0$
and its conjugate (as is the case if $J$ is
irreducible, as shown in Table \ref{T:superforms}), then $J_0$ will be a chiral
superfield and the highest-dimension component of $J$ will be
$J_{abcd}=\frac{i}{8}\e_{abcd}(D^2J_0-\bar{D}^2\bar{J}_0)$.
Therefore, the action \eqref{E:EctoAction} takes the manifestly supersymmetric form
\be
S_J=\operatorname{Re}\left[ i\int d^4xd^2\t J_0 \right]\!.
\ee

Note that $S_{dL}=0$ for any globally defined 3-form $L$.
We can use this as follows:
As pointed out in section \ref{S:pForms}, if the closed super-4-form $J$ is obtained from lower-degree forms by wedging, $J_{\dot\al\dot\beta ab}$ will generally not be the lowest-dimensional component.
Then we can try to shift $J$ by an exact form, $J'=J-dL$ (for some gauge-invariant 3-form $L$),
so that the lowest dimensional component of $J'$ is $J'_{\dot\al\dot\beta ab}$. In this case,
\be
S_J=S_{J'}=\operatorname{Re}\left[ i\int d^4xd^2\t J'_0 \right]\!.
\ee
We now extend this construction to the non-abelian hierarchy.

The ectoplasmic invariants for the non-abelian tensor hierarchy are the natural analogs of (\ref{E:EctoAction}) suggested by the substitution $d \to \mathcal Q$.
More specifically, the condition is that if $\{J^{(p)}\}_{p=0}^D$ is a collection of superforms with bi-degree $(p,D-p)$ and with $(\Q J)^{(5)}=0$, then
\be
S_J=\tfrac{1}{24}\int d^4x\int_M\e^{abcd}J^{(4)}_{abcd}\big|
\ee
is a supersymmetric action. Moreover, if the lowest-dimension component of $J^{(4)}$ is $J^{(4)}_{\dot\al\dot\beta ab}=-4(\bar{\s}_{ab})_{\dot\al\dot\beta}J_0$ then
\be
\bar{\D}_{\dot\al}J_0=0,\quad S_J=\operatorname{Re}\left[ i\int d^4xd^2\t\int_MJ_0\right]\!.
\ee
The proofs of these statements follow the same steps as above, but with some extra terms getting dropped.  For instance,
\begin{align}
\d S_J & = \tfrac{1}{24} \bar{\varepsilon}^{\dot\al}\int d^4x\int_M\e^{abcd} \bar{Q}_{\dot\al}J^{(4)}_{abcd} \big|= \tfrac{1}{24}\bar{\varepsilon}^{\dot\al}\int d^4x\int_M\e^{abcd}\bar{D}_{\dot\al}J^{(4)}_{abcd} \big|\non\\
&= \tfrac{1}{24} \bar{\varepsilon}^{\dot\al}\int d^4x\int_M\e^{abcd}( 4\D_aJ^{(4)}_{\dot\al bcd}+\p J^{(5)}_{\dot\al abcd}+4(\iota_\mcF)_{\dot\al a}J^{(3)}_{bcd})\big|=0.
\end{align}
Here the first term vanishes because it is a combination of a total spacetime derivative and a piece that is a Lie derivative of a top form on $M$.
The second term vanishes because it is a total derivative on $M$. 
Finally, the last form $J^{(3)}$ must be zero because it is a bosonic $(D-3)$-form on a $(D-4)$-dimensional manifold.  The other proofs proceed similarly.
By using the same manipulations, we can also show that if the polyform $J$ is $\Q$-exact then $S_J=0$.

\subsection{Weil Triviality}
\label{S:WeilTriviality}

Our interest is in supersymmetrizing Chern-Simons actions.  The basic bosonic action can be given by defining a Chern-Simons super(-poly)form using the potential and field-strength superforms $C=\sum_pC^{(p)}$ and $F=\sum_pF^{(p)}$,
\be
\om_n=C\w F^{n-1},
\ee
and then integrating $\om_n^{(4)}\big|$ over four-dimensional spacetime and the internal space,
\be
S_{CS,n}= \tfrac{1}{24}\int d^4x\e^{abcd}\int_M\om^{(4)}_{n\,abcd}\big|.
\ee
This action is gauge invariant under both the non-abelian gauge transformations (since it is invariant under internal diffeomorphisms by construction) and under the abelian gauge transformations which leave $F$ invariant and transform $C$ by $\d C=\Q\Lambda$. To see this,
\be
\d S_{CS,n}=\tfrac{1}{24}\int d^4x\e^{abcd}\int_M(\Q(\Lambda\w F^{n-1}))_{abcd}\big|=0,
\ee
where we exploit the fact that the integral of a $\Q$-exact form is zero from integration by parts on spacetime and on the internal manifold, and the fact that a top (bosonic) form cannot be the contraction of anything.

Of course $\om_n$ is not $\Q$-closed in general, and in particular $(\Q\om_n)^{(5)}\ne 0$ except in the trivial case of $n=1$, so $S_{CS,n}$ is not supersymmetric in general.  The case $n=1$ is special and does give a supersymmetric result
\be
S_{SCS,1}=S_{CS,1}=\tfrac{1}{24}\int d^4x\e^{abcd}\int_MC^{(4)}_{abcd}\big|=\Real\left[ i\int d^4xd^2\t\int_M\G\right]\!.
\ee
For $n>1$, the act of supersymmetrization involves finding a different polyform $K_n$ that is invariant under abelian gauge transformations and satisfies $\Q K_n=F^n=\Q\om_n$. (Actually we only need it to hold in degree five, $(\Q K_n)^{(5)}=(F^n)^{(5)}$.) In this case the form $J_n=\om_n-K_n$ is $\Q$-closed and can be used to build a supersymmetric, gauge-invariant action.
The existence of the gauge-invariant class $K$ is a phenomenon known as Weil triviality \cite{Bonora:1986xd}.

We will use (relative \cite{Howe:2011tm}) cohomology \cite{Howe:1998tsa, Berkovits:2008qw} to construct the relevant superinvariants.
This was applied to the construction of Chern-Simons-like invariants in \cite{Butter:2012ze, Kuzenko:2013rna, Kuzenko:2014jra}
based on earlier work on Weil triviality \cite{Bonora:1986xd}.
In the next section we will implement this procedure in the case of the quadratic and higher-order Chern-Simons actions.

\section{All Chern-Simons Actions}
\label{S:AllCS}

In this section we prove Claim \ref{Th:Recursive} in subsection \ref{S:Proof} and use it to find all possible Chern-Simons-like invariants of the form (\ref{eq:CompositeAction}) in subsection \ref{S:Solution}. 
The two subsections are independent and so the reader interested only in the solution can skip directly to subsection \ref{S:Solution}. 

We begin by dispensing with an ambiguity in the formalism of section \ref{S:PrepotentialNATH}: Suppose we have another set of globally defined (so built from the basic field-strengths $E$, $U$, $W^\al$, $H$, and $G$) composite superfields $\{\phi,v,\s^\al,x,\g\}$.  Then we can always get a solution to (\ref{eqs:DescentRelations}) by constructing $\{e,u,w^\al,h,g\}$ from $\{\phi,v,\s^\al,x,\g\}$ in the same way that the set $\{E,U,W^\al,H,G\}$ is constructed from $\{\Phi,V,\Sigma^\al,X,\G\}$ in (\ref{eqs:SuperfieldStrengths}). That is, the composite field-strengths would be exact in the \textit{field-strengths}, not the prepotentials. The corresponding action (\ref{eq:CompositeAction}) can then be written, after integrations by parts, in terms of field-strengths alone.  So if we are interested in super-Chern-Simons actions which are not equivalent to completely gauge-invariant constructions, then such actions are trivial.  This is consistent with the cohomological formulation of Chern-Simons actions presented in \cite{Becker:2016rku}.

Moreover, an ``exact" (in the cohomological sense) composite action of this sort can even give zero contribution to the action.  For constructions involving dimensional reduction, where we expect the composite fields to be internal forms of the appropriate degree and we restrict the engineering dimension to match the standard Chern-Simons action, this can only happen in contributions to $x$. In particular, we can have
\begin{align}
x &= \sum_{k=0}^{n-3}a_kE^k\bar{E}^{n-3-k}UH +\sum_{k=0}^{n-4}( b_kE^k\bar{E}^{n-4-k}U\D^\al UW_\al+\mathrm{c.c.})\non\\
&\quad +\sum_{k=0}^{n-5}(c_kE^k\bar{E}^{n-5-k}U\D^\al U\bar{\D}_{\dot\al}U\D_\al\bar{\D}^{\dot\al}U+\mathrm{c.c.}),
\end{align}
with $a_k$, $b_k$, and $c_k$ being complex constants (and with $\bar{a}_k=a_{n-3-k}$).  Constructing the corresponding $h=-\p x$ and $g=-\tfrac{1}{4}\bar{\D}^2x$ and plugging into the action (\ref{eq:CompositeAction}) gives zero after integration by parts (it will be proportional to the wedge product of two $U$'s, which vanishes since $U$ is an odd degree form on the internal space). We must keep this ambiguity in mind when constructing solutions to (\ref{eqs:DescentRelations}).

\subsection{Proof of Claim \ref{Th:Recursive}}
\label{S:Proof}

As mentioned below Claim \ref{Th:Recursive}, the most obvious way to verify the stated recursion relations is to plug them into the Bianchi identities and check that they are satisfied identically. Doing so would prove the claim but does not give any insight into how the recursion relation is found or why a solution may be expected to exist in the first place, so we have decided to present a constructive proof in terms of the supergeometry underlying the non-abelian tensor hierarchy instead. The proof of the all-orders action (\S \ref{S:AllOrders}) is longer and more technical than that for the quadratic action (\S \ref{S:Quadratic}) so we present the latter first.

\subsubsection{Quadratic Action}
\label{S:Quadratic}

We need to find a polyform $K_2$ which is gauge invariant (so only constructed out of field-strength superfields), and that satisfies $(\Q K_2)^{(5)}=(F^2)^{(5)}$.  Because of the way the $\Q$ operator mixes different degrees, we must proceed systematically.  Define the {\it weight} of a superform component to be the sum of the number of vector and un-dotted spinor indices.  Then any component of $\Q K_2$ of weight $w$ only depends on components of $K_2$ whose weights are less than or equal to $w$ (cf.\ eq.\ \ref{E:Jank}). For this reason, we can build $K_2$ starting with the lowest weights.  Within a given weight, we start at the highest degree and work down. The components of the superforms used here are reviewed and defined in appendix \ref{S:MoreForms}.

The lowest-weight component of $(F^2)^{(5)}$ is
\be
( F^2)^{(5)}_{\bar{s}\bar{s}\bar{s}\psi\psi}=6F^{(1)}_{\bar{s}}F^{(4)}_{\bar{s}\bar{s}\psi\psi}=24i(\bar{\s}_{\psi\psi})_{\bar{s}\bar{s}}\bar{\D}_{\bar{s}}UG.
\ee
Comparing with\footnote{We can assume that $K_2$ has no components of weight zero above degree two, weight one above degree three, or weight two above degree four since they are not required to match any non-zero components of $(F^2)^{(5)}$.}
\be
(\Q K_2)^{(5)}_{\bar{s}\bar{s}\bar{s}\psi\psi}=3\bar{\D}_{\bar{s}}K^{(4)}_{2\,\bar{s}\bar{s}\psi\psi},
\ee
we deduce that we can set
\be
K^{(4)}_{2\,\bar{s}\bar{s}\psi\psi}=8i(\bar{\s}_{\psi\psi})_{\bar{s}\bar{s}}UG.
\ee
We now have the relevant components of a polyform $J_2=\om_2-K_2$ satisfying $(\Q J_2)^{(5)}=0$, but we still need to put it in the form where we can read off the action. (The remaining, higher-weight components of $K_2$ necessary to fully fix $(\Q J_2)^{(5)} = 0$ do not enter into the following analysis so we will not bother to present them here.) To do this, we must remove the component $J^{(4)}_{2\,s\bar{s}\bar{s}\psi}$ by subtracting an exact piece $(\Q L_2)^{(4)}$ which will not affect the action. We have
\be
J^{(4)}_{2\,s\bar{s}\bar{s}\psi}=-2C^{(1)}_{\bar{s}}F^{(3)}_{s\bar{s}\psi}+2C^{(3)}_{s\bar{s}\psi}F^{(1)}_{\bar{s}}=-2(\s_\psi)_{s\bar{s}}\bar{\D}_{\bar{s}}VH+2(\s_\psi)_{s\bar{s}}X\bar{\D}_{\bar{s}}U.
\ee
We can remove this by choosing a polyform $L_2$ with no weight zero or one components, and whose first weight two component is
\be
L^{(3)}_{2\,\bar{s}\psi\psi}=-i(\bar{\s}_{\psi\psi})^{\dot\al}_{\hph{\dot\al}\bar{s}}(\bar{\D}_{\dot\al}VH-X\bar{\D}_{\dot\al}U).
\ee
Then the action can be read off from the component
\begin{align}
( J_2-\Q L_2)^{(4)}_{\bar{s}\bar{s}\psi\psi} &=C^{(0)}F^{(4)}_{\bar{s}\bar{s}\psi\psi}-2C^{(1)}_{\bar{s}}F^{(3)}_{\bar{s}\psi\psi}-8C^{(2)}_{\bar{s}\psi}F^{(2)}_{\bar{s}\psi}-2C^{(3)}_{\bar{s}\psi\psi}F^{(1)}_{\bar{s}}\non\\
& \quad +C^{(4)}_{\bar{s}\bar{s}\psi\psi}F^{(0)}-K^{(4)}_{2\,\bar{s}\bar{s}\psi\psi}-2\bar{\D}_{\bar{s}}L^{(3)}_{\bar{s}\psi\psi}\non\\
&=-4(\bar{\s}_{\psi\psi})_{\bar{s}\bar{s}}\left[\Phi G+\Sigma^\al W_\al+\G E+\frac{i}{4}\bar{\D}^2( VH-XU)\right].
\end{align}
Note that all explicit $\partial$'s and $\iota$'s have canceled out of this expression.

This leads to the action
\be
S=\Real\left[ i\int d^4xd^2\t\int_M(\Phi G+\Sigma^\al W_\al+\G E)\right]+\int d^4xd^4\t\int_M( VH-XU).
\ee

\subsubsection{Higher-order Actions}
\label{S:AllOrders}

Now we will show that the procedure above can be adapted for higher orders in the number of fields as well.  By our previous arguments and the structure of $\Q$-cohomology, it will always be possible to find polyforms $K_n$ and $L_n$ such that $J_n=\om_n-K_n$ satisfies $(\Q J_n)^{(5)}=0$ and $(J_n-\Q L_n)^{(4)}$ has $(J_n-\Q L_n)^{(4)}_{\bar{s}\bar{s}\psi\psi}=-4(\bar{\s}_{\psi\psi})_{\bar{s}\bar{s}}J_{n\,0}$ as its lowest non-vanishing component.  In practice however, this can be very computationally intensive for $n>3$ and is a significant calculation even for $n=3$.  However, we will argue that there is a shortcut.  Suppose one can find a gauge-invariant polyform $M_n$ such that $\mathbf{F}_n=F^{n-1}-\Q M_n$ has the same components as $F$. That is,
\begin{itemize}
\item the only weight zero components of $\mathbf{F}_n$ are $\mathbf{F}^{(1)}_{n\,\bar{s}}=-i\bar{\D}u_n$ and $\mathbf{F}^{(0)}_n=\hlf(e_n+\bar{e}_n)$, where $u_n$ is a real superfield and $e_n$ is chiral;
\item the only weight one component of $\mathbf{F}_n$ above degree one is $\mathbf{F}^{(2)}_{n\,\bar{s}\psi}=-(\s_\psi)_{\al\bar{s}}(w_n^\al-i(\iota_\mcW)^\al u_n)$, where $w_n^\al$ is chiral;
\item the only weight two components of $\mathbf{F}_n$ above degree two are $\mathbf{F}^{(4)}_{n\,\bar{s}\bar{s}\psi\psi}=-4(\bar{\s}_{\psi\psi})_{\bar{s}\bar{s}}g_n$, $\mathbf{F}^{(3)}_{n\,s\bar{s}\psi}=i(\s_\psi)_{s\bar{s}}h_n$, and $\mathbf{F}^{(3)}_{n\,\bar{s}\psi\psi}=-(\bar{\s}_{\psi\psi})^{\dot\al}_{\hph{\dot\al}\bar{s}}\bar{\D}_{\dot\al}h_n$, where $g_n$ is chiral and $h_n$ is real;
\item and the superfields $e_n$, $u_n$, $w_n^\al$, $h_n$, and $g_n$ are gauge-invariant composites, constructed from (and of degree $n-1$ in) the field-strengths $\{E, U, W^\al, H, G\}$.
\end{itemize}
If we can find such an $M_n$, then we have
\be
\om_n=C\w\mathbf{F}_n-\Q( C\w M_n)+F\w M_n.
\ee
The third term above is already completely gauge invariant and corresponds to adding a piece to the action which can be written purely in terms of field-strengths (and hence represents an ambiguity in the super-Chern-Simons action).  The second term is $\Q$-exact and hence will not contribute to the action. Thus, we are free to replace $\om_n$ by $C\w\mathbf{F}_n$ in our construction of $J_n$.  Once we have done that, the procedure to find the action proceeds exactly as in the quadratic case.  In particular, we need to find $K_n$ such that $(\Q K_n)^{(5)}=F\w \mathbf{F}_n$. The only component we need comes from
\be
( F\w\mathbf{F}_n)^{(5)}_{\bar{s}\bar{s}\bar{s}\psi\psi}=3F^{(1)}_{\bar{s}}\mathbf{F}^{(4)}_{n\,\bar{s}\bar{s}\psi\psi}+3F^{(4)}_{\bar{s}\bar{s}\psi\psi}\mathbf{F}^{(1)}_{n\,\bar{s}}=12i(\bar{\s}_{\psi\psi})_{\bar{s}\bar{s}}(\bar{\D}_{\bar{s}}Ug_n+G\bar{\D}_{\bar{s}}u_n),
\ee
leading to
\be
K^{(4)}_{n\,\bar{s}\bar{s}\psi\psi}=4i(\bar{\s}_{\psi\psi})_{\bar{s}\bar{s}}( Ug_n+Gu_n).
\ee
To get rid of
\be
J^{(4)}_{n\,s\bar{s}\bar{s}\psi}=-2C^{(1)}_{\bar{s}}\mathbf{F}^{(3)}_{n\,s\bar{s}\psi}+2C^{(3)}_{s\bar{s}\psi}\mathbf{F}^{(1)}_{n\,\bar{s}}=-2(\s_\psi)_{s\bar{s}}(\bar{\D}_{\bar{s}}Vh_n-X\bar{\D}_{\bar{s}}u_n),
\ee
we set
\be
L^{(3)}_{n\,\bar{s}\psi\psi}=-i(\bar{\s}_{\psi\psi})^{\dot\al}_{\hph{\dot\al}\bar{s}}(\bar{\D}_{\dot\al}Vh_n-X\bar{\D}_{\dot\al}u_n).
\ee
Putting the pieces together, this gives
\be
J_{n\,0}=\Phi g_n+\Sigma^\al w_{n\,\al}+\G e_n+\frac{i}{4}\bar{\D}^2( Vh_n-Xu_n),
\ee
and
\begin{align}
S_{SCS,n}& =\Real\left[ i\int d^4xd^2\t\int _M(\Phi g_n+\Sigma^\al w_{n\,\al}+\G e_n)\right]\notag\\
& \quad +\int d^4xd^4\t\int_M( Vh_n-Xu_n).
\end{align}

What remains is to find $M_n$ and the components of $\mathbf{F}_n$.  Suppose that we have already found the solution for $n-1$.  We need $M_n$ to satisfy
\be
\Q M_n=F^{n-1}-\mathbf{F}_n=F\w F^{n-2}-\mathbf{F}_n=F\w\mathbf{F}_{n-1}+\Q( F\w M_{n-1})-\mathbf{F}_n,
\ee
so we can set $M_n= F\wedge M_{n-1}+\d M$, where $\d M$ is such that $F\w\mathbf{F}_{n-1}-\Q(\d M)$ has the same components as $F$.  But this is a short task since both $F$ and $\mathbf{F}_{n-1}$ have been put into the same simple form and so the procedure to find $\d M$ is the same as finding $K_2$, though here we need to find more components.  As we construct $\d M$, we can also read off the components of $\mathbf{F}_n=F\w\mathbf{F}_{n-1}-\Q\d M$.

Starting with weight zero, we have at degree two
\be
( F\w\mathbf{F}_{n-1})^{(2)}_{\bar{s}\bar{s}}=2F^{(1)}_{\bar{s}}\mathbf{F}^{(1)}_{n-1\,\bar{s}}=2\bar{\D}_{\bar{s}}U\bar{\D}_{\bar{s}}u_{n-1},
\ee
leading to
\be
\d M^{(1)}_{\bar{s}}=U\bar{\D}_{\bar{s}}u_{n-1}.
\ee
At degree one we have
\begin{align}
( F\w\mathbf{F}_{n-1}-\Q\d M)^{(1)}_{\bar{s}} &=F^{(0)}\mathbf{F}^{(1)}_{n-1\,\bar{s}}+F^{(1)}_{\bar{s}}\mathbf{F}_{n-1}^{(0)}+\p\d M^{(1)}_{\bar{s}}\non\\
&=-\frac{i}{2}( E+\bar{E})\bar{\D}_{\bar{s}}u_{n-1}-\frac{i}{2}\bar{\D}_{\bar{s}}U( e_{n-1}+\bar{e}_{n-1})\non\\
& \quad +\p U\bar{\D}_{\bar{s}}u_{n-1}-U\bar{\D}_{\bar{s}}\p u_{n-1}\non\\
&=-iE\bar{\D}_{\bar{s}}u_{n-1}-\frac{i}{2}\bar{\D}_{\bar{s}}Ue_{n-1}-\frac{i}{2}\bar{\D}_{\bar{s}}U\bar{e}_{n-1}-\frac{i}{2}U\bar{\D}_{\bar{s}}\bar{e}_{n-1}\non\\
&=-i\bar{\D}_{\bar{s}}u_n+\bar{\D}_{\bar{s}}\d M^{(0)},
\end{align}
where
\be
u_n=\hlf( E+\bar{E}) u_{n-1}+\hlf U( e_{n-1}+\bar{e}_{n-1}),\quad \d M^{(0)}=-\frac{i}{2}( E-\bar{E}) u_{n-1},
\ee
and where we made use of the Bianchi identities
\be
\p U=-\frac{i}{2}( E-\bar{E}),\quad\p u_{n-1}=-\frac{i}{2}( e_{n-1}-\bar{e}_{n-1}),
\ee
which are simply a consequence of the $\Q$-closure of $F$ and $\mathbf{F}_{n-1}$, respectively.  The final weight zero piece is at degree zero,
\begin{align}
( F\w\mathbf{F}_{n-1}-\Q M)^{(0)} &=F^{(0)}\mathbf{F}^{(0)}_{n-1}-\p\d M^{(0)}\non\\
&=\frac{1}{4}( E+\bar{E})( e_{n-1}+\bar{e}_{n-1})+\frac{i}{2}( E-\bar{E})\p u_{n-1}\non\\
& =\hlf( e_n+\bar{e}_n),
\end{align}
where
\be
e_n=Ee_{n-1}.
\ee
Note that we have now derived the recursion relations (\ref{eqs:Recursion}) for $e_n$ and $u_n$. Proceeding with weight one, we have at degree three,
\begin{align}
( F\w\mathbf{F}_{n-1}-\Q\d M)^{(3)}_{\bar{s}\bar{s}\psi} &=2F^{(1)}_{\bar{s}}\mathbf{F}^{(2)}_{n-1\,\bar{s}\psi}-2F^{(2)}_{\bar{s}\psi}\mathbf{F}^{(1)}_{n-1\,\bar{s}}-2(\iota_\mcF)_{\bar{s}\psi}\d M^{(1)}_{\bar{s}}\non\\
&=2i(\s_\psi)_{\al\bar{s}}(\bar{\D}_{\bar{s}}U( w_{n-1}^\al-i(\iota_\mcW)^\al u_{n-1})\non\\
& \quad -( W^\al-i(\iota_\mcW)^\al U)\bar{\D}_{\bar{s}}u_{n-1})\non\\
& \quad +2(\s_\psi)_{\al\bar{s}}(\iota_\mcW)^\al( U\bar{\D}_{\bar{s}}u_{n-1})\non\\
&=2i(\s_\psi)_{\al\bar{s}}(\bar{\D}_{\bar{s}}Uw_{n-1}^\al-W^\al\bar{\D}_{\bar{s}}u_{n-1}\non\\
& \quad -i\bar{\D}_{\bar{s}}U(\iota_\mcW)^\al u_{n-1}-iU\bar{\D}_{\bar{s}}(\iota_\mcW)^\al u_{n-1}),
\end{align}
which leads to
\be
\d M^{(2)}_{\bar{s}\psi}=i(\s_\psi)_{\al\bar{s}}( Uw_{n-1}^\al+W^\al u_{n-1}-iU(\iota_\mcW)^\al u_{n-1}).
\ee
At degree two---recalling that $M^{(1)}_s$ is fixed to be the conjugate of $M^{(1)}_{\bar{s}}$ since all of our superforms are real---
\begin{align}
( F\w\mathbf{F}_{n-1}-\Q\d M)^{(2)}_{s\bar{s}} &=-F^{(1)}_s\mathbf{F}^{(1)}_{n-1\,\bar{s}}-F^{(1)}_{\bar{s}}\mathbf{F}^{(1)}_{n-1\,s}-\D_s\d M^{(1)}_{\bar{s}}-\bar{\D}_{\bar{s}}\d M^{(1)}_s\non\\
&=-\D_sU\bar{\D}_{\bar{s}}u_{n-1}-\bar{\D}_{\bar{s}}U\D_su_{n-1}\non\\
& \quad -\D_s( U\bar{\D}_{\bar{s}}u_{n-1})-\bar{\D}_{\bar{s}}( U\D_su_{n-1}),
\end{align}
which can be canceled by choosing
\be
\d M^{(1)}_\psi=-\frac{i}{2}(\s_\psi)_{\al\dot\al}(\D^\al U\bar{\D}^{\dot\al}u_{n-1}+\bar{\D}^{\dot\al}U\D^\al u_{n-1}+\hlf U\{\D^\al,\bar{\D}^{\dot\al}\}u_{n-1}).
\ee
We can then read off $w_n^\al$ from
\begin{align}
( F\w\mathbf{F}_{n-1}-\Q\d M)^{(2)}_{\bar{s}\psi} &=F^{(0)}\mathbf{F}^{(2)}_{n-1\,\bar{s}\psi}+F^{(1)}_{\bar{s}}\mathbf{F}^{(1)}_{n-1\,\psi}-F^{(1)}_\psi\mathbf{F}^{(1)}_{\bar{s}}+F^{(2)}_{\bar{s}\psi}\mathbf{F}_{n-1}^{(0)}\non\\
& \quad -\bar{\D}_{\bar{s}}\d M^{(1)}_\psi+\D_\psi\d M^{(1)}_{\bar{s}}-\p\d M^{(2)}_{\bar{s}\psi}-(\iota_\mcF)_{\bar{s}\psi}\d M^{(0)}\non\\&=:-i(\s_\psi)_{\al\bar{s}}( w_n^\al-i(\iota_\mcW)^\al u_n),
\end{align}
which gives (after some algebra)
\be
w_n^\al=Ew_{n-1}^\al+W^\al e_{n-1}+\frac{i}{4}\bar{\D}^2(\D^\al Uu_{n-1}-U\D^\al u_{n-1}).
\ee
Moving on to weight two, we have at degree five,
\begin{align}
( F\w\mathbf{F}_{n-1})^{(5)}_{\bar{s}\bar{s}\bar{s}\psi\psi} &=3F^{(1)}_{\bar{s}}\mathbf{F}^{(4)}_{n-1\,\bar{s}\bar{s}\psi}+3F^{(4)}_{\bar{s}\bar{s}\psi\psi}\mathbf{F}^{(1)}_{\bar{s}}\non\\
&=12i(\bar{\s}_{\psi\psi})_{\bar{s}\bar{s}}(\bar{\D}_{\bar{s}}Ug_{n-1}+G\bar{\D}_{\bar{s}}u_{n-1}),
\end{align}
which can be canceled by setting
\be
\d M^{(4)}_{\bar{s}\bar{s}\psi\psi}=4i(\bar{\s}_{\psi\psi})_{\bar{s}\bar{s}}( Ug_{n-1}+Gu_{n-1}).
\ee
Going down to degree four, we must first cancel
\begin{align}
( F\w\mathbf{F}_{n-1})^{(4)}_{s\bar{s}\bar{s}\psi} &=-F^{(1)}_{\bar{s}}\mathbf{F}^{(3)}_{n-1\,s\bar{s}\psi}+F^{(3)}_{s\bar{s}\psi}\mathbf{F}^{(1)}_{n-1\,\bar{s}}\non\\
&=(\s_\psi)_{s\bar{s}}( -\bar{\D}_{\bar{s}}Uh_{n-1}+H\bar{\D}_{\bar{s}}u_{n-1}),
\end{align}
which can be done by setting
\be
\d M^{(3)}_{\bar{s}\psi\psi}=i(\bar{\s}_{\psi\psi})^{\dot\al}_{\hph{\dot\al}\bar{s}}(\bar{\D}_{\dot\al}Uh_{n-1}-H\bar{\D}_{\dot\al}u_{n-1}).
\ee
Then we can read off $g_n$ from
\begin{align}
( F\w\mathbf{F}_{n-1}-\Q\d M)^{(4)}_{\bar{s}\bar{s}\psi\psi} &=F^{(0)}\mathbf{F}^{(4)}_{n-1\,\bar{s}\bar{s}\psi\psi}-2F^{(1)}_{\bar{s}}\mathbf{F}^{(3)}_{n-1\,\bar{s}\psi\psi}-4F^{(2)}_{\bar{s}\psi}\mathbf{F}^{(2)}_{n-1\,\bar{s}\psi}\non\\
& \quad -F^{(3)}_{\bar{s}\psi\psi}\mathbf{F}^{(1)}_{n-1\,\bar{s}}+F^{(4)}_{\bar{s}\bar{s}\psi\psi}\mathbf{F}^{(0)}-2\bar{\D}_{\bar{s}}\d M^{(3)}_{\bar{s}\psi\psi}\non\\
& \quad -\p\d M^{(4)}_{\bar{s}\bar{s}\psi\psi}+(\iota_\mcF)_{\bar{s}\psi}\d M^{(2)}_{\bar{s}\psi}\non\\
& \eqc -4(\bar{\s}_{\psi\psi})_{\bar{s}\bar{s}}g_n,
\end{align}
where, again suppressing some algebra,
\be
g_n=Eg_{n-1}+Ge_{n-1}+W^\al w_{n-1\,\al}+\frac{i}{4}\bar{\D}^2( Hu_{n-1}-Uh_{n-1}).
\ee
Finally, we have
\begin{align}
( F\w\mathbf{F}_{n-1}-\Q M)^{(3)}_{s\bar{s}\psi} &=F^{(0)}\mathbf{F}^{(3)}_{n-1\,s\bar{s}\psi}+F^{(1)}_s\mathbf{F}^{(2)}_{\bar{s}\psi}+F^{(1)}_{\bar{s}}\mathbf{F}^{(2)}_{s\psi}-2F^{(2)}_{s\psi}\mathbf{F}^{(1)}_{\bar{s}}\non\\
& \quad -2F^{(2)}_{\bar{s}\psi}\mathbf{F}^{(1)}_s+F^{(3)}_{s\bar{s}\psi}\mathbf{F}^{(0)}-\D_s\d M^{(2)}_{\bar{s}\psi}-\bar{\D}_{\bar{s}}\d M^{(2)}_{s\psi}\non\\
& \quad -(\iota_\mcF)_{s\psi}\d M^{(1)}_{\bar{s}}-(\iota_\mcF)_{\bar{s}\psi}\d M^{(1)}_s\non\\
& \eqc i(\s_\psi)_{s\bar{s}}h_n+2i\s^a_{s\bar{s}}\d M^{(2)}_{\psi a}.
\end{align}
We do not have to directly compute $\d M^{(2)}_{\psi\psi}$ because $h_n$ can be isolated by contracting, yielding $h_n=\frac{i}{8}(\bar{\s}^a)^{\dot\al\al}(F\w\mathbf{F}_{n-1}-\Q M)^{(3)}_{\al\dot\al a}$. This leads to
\begin{align}
h_n &=\hlf( E+\bar{E}) h_{n-1}+\hlf H( e_{n-1}+\bar{e}_{n-1})+\Om(w_{n-1},U)+\Om(W,u_{n-1})\non\\
& \quad -i\D^\al U(\iota_\mcW)_\al u_{n-1}-i(\iota_\mcW)^\al U\D_\al u_{n-1} \notag\\
& \quad +i\bar{\D}_{\dot\al}U(\iota_{\bar{\mcW}})^{\dot\al}u_{n-1}+i(\iota_{\bar{\mcW}})_{\dot\al}U\bar{\D}^{\dot\al}u_{n-1}
\end{align}
and completes the derivation of the recursion relations \eqref{eqs:Recursion}.

\subsection{Solution of the Recursion Relations}
\label{S:Solution}
In this section, we solve the recursion relations \eqref{eqs:Recursion} to all orders in $n$.  To do this, we will first treat $n$ as a continuous parameter.  Then the first recursion relation can be written as
\be
( e^{\frac{d}{dn}}-E) e_n=0,
\ee
which has the general solution
\be
e_n=c E^n
\ee
for some $n$-independent quantity $c$.  Since we want $e_2=E$, this fixes $c=E^{-1}$ and the general solution is $e_n=E^{n-1}$.

Moving on to the next equation we have
\be
\left( e^{\frac{d}{dn}}-\frac{E+\bar{E}}{2}\right) u_n=\hlf( E^{n-1}+\bar{E}^{n-1}) U.
\ee
In general, for constants $a$ and $b$ we have
\be
( e^{\frac{d}{dn}}-a)^{-1}[ b^n]=\left\{\begin{matrix}\frac{b^n}{b-a}, & \mathrm{if\ }b\ne a, \\ na^{n-1}, & \mathrm{if\ }b=a.\end{matrix}\right.
\ee
Then we can find the solution for $u_n$ by adding the homogeneous solution to a particular solution,
\bea
u_n &=& c \left(\frac{E+\bar{E}}{2}\right)^n+\hlf\left( e^{\frac{d}{dn}}-\frac{E+\bar{E}}{2}\right)^{-1}[( E^{n-1}+\bar{E}^{n-1}) U]\non\\
&=& c\left(\frac{E+\bar{E}}{2}\right)^n+\frac{E^{n-1}-\bar{E}^{n-1}}{E-\bar{E}}U.
\eea
Demanding that $u_2=U$ fixes $c=0$, so
\be
u_n=\frac{E^{n-1}-\bar{E}^{n-1}}{E-\bar{E}}U.
\ee
Note that the denominator can always be canceled, so the solution is always polynomial in the fields.

Proceeding, we have
\bea
w_n^\al &=& c^\al E^n+( e^{\frac{d}{dn}}-E)^{-1}\left[ E^{n-1}W^\al-\frac{i}{2}\bar{\D}^2\left(\frac{E^{n-1}-\bar{E}^{n-1}}{E-\bar{E}}U\D^\al U\right)\right]\non\\
&=& c^\al E^n+nE^{n-2}W^\al-\frac{i}{2}\bar{\D}^2\left[\left(\frac{nE^{n-2}}{E-\bar{E}}+\frac{\bar{E}^{n-1}}{( E-\bar{E})^2}\right) U\D^\al U\right]\!.
\eea
Matching $n=2$ requires
\be
c^\al=-E^{-2}W^\al+\frac{i}{2}E^{-2}\bar{\D}^2\left(\frac{2E-\bar{E}}{( E-\bar{E})^2}U\D^\al U\right)\!,
\ee
giving
\begin{align}
w_n^\al =& ( n-1) E^{n-2}W^\al-\frac{i}{2}\bar{\D}^2\left[\left(\frac{( n-1) E^{n-2}}{E-\bar{E}}-\frac{E^{n-1}-\bar{E}^{n-1}}{( E-\bar{E})^2}\right) UD^\al U\right]\non\\
& =( n-1) E^{n-2}W^\al-\frac{i}{2}\bar{\D}^2(\e_nU\D^\al U),
\end{align}
where
\be
\e_n=\frac{\p}{\p E}\frac{u_n}{U}=\frac{( n-1) E^{n-2}}{E-\bar{E}}-\frac{E^{n-1}-\bar{E}^{n-1}}{( E-\bar{E})^2}.
\ee

The solutions for $h_n$ and $g_n$ can be obtained similarly, but the expressions are unilluminating.  Instead, we will simply list the results for $n=3$ and $n=4$.
The cubic Chern-Simons invariant is defined by the composite superfields
\begin{subequations}
\label{eqs:n3sols}
\begin{align}
e_3 &= E^2,\\
u_3 &= ( E+\bar{E}) U,\\
w_3^\al &= 2EW^\al-\frac{i}{2}\bar{\D}^2( U\D^\al U),\\
h_3 &= ( E+\bar{E}) H+2\Om(W,U)-2i\D^\al U(\iota_\mcW)_\al U+2i\bar{\D}_{\dot\al}U(\iota_{\bar{\mcW}})^{\dot\al}U,\\
g_3 &= 2EG+W^\al W_\al-\frac{i}{2}\bar{\D}^2( UH),
\end{align}
\end{subequations}
The solution agrees with the results found in \cite{Becker:2016rku}.
The quartic Chern-Simons invariant is defined by the following solution to the descent relations:
\begin{subequations}
\label{eqs:n4sols}
\begin{align}
e_4 &= E^3,\\
u_4 &= ( E^2+E\bar{E}+\bar{E}^2) U,\\
w_4^\al &= 3EW^\al-\frac{i}{2}\bar{\D}^2[( 2E+\bar{E}) U\D^\al U],\notag\\
h_4 &= ( E^2+E\bar{E}+\bar{E}^2) H\non\\
& \quad +[( 2E+\bar{E})( 2\D^\al UW_\al+U\D^\al W_\al)+2\D^\al EUW_\al+\mathrm{c.c.}]\non\\
& \quad +\frac{i}{4}U\D^\al U(\bar{\D}^2\D_\al U-2\bar{\D}_{\dot\al}\D_\al\bar{\D}^{\dot\al}U+\D_\al\bar{\D}^2U)\non\\
& \quad -\frac{i}{4}U\bar{\D}_{\dot\al}U(\D^2\bar{\D}^{\dot\al}U-2\D^\al\bar{\D}^{\dot\al}\D_\al U+\bar{\D}^{\dot\al}\D^2U)-iU\D^\al\bar{\D}_{\dot\al}U\bar{\D}^{\dot\al}\D_\al U\non\\
& \quad +\frac{i}{2}U\D^2U\bar{\D}^2U+i\D^\al U\bar{\D}_{\dot\al}U\D^\al\bar{\D}_{\dot\al}U-i\D^\al U\bar{\D}_{\dot\al}U\bar{\D}^{\dot\al}\D_\al U\non\\
& \quad -2i( E+\bar{E})(\D^\al U(\iota_\mcW)_\al U-\bar{\D}_{\dot\al}U(\iota_{\bar{\mcW}})^{\dot\al}U)+i\D^\al EU(\iota_\mcW)_\al U\non\\
& \quad -i\bar{\D}_{\dot\al}\bar{E}U(\iota_{\bar{\mcW}})^{\dot\al}U-i(\iota_\mcW)^\al( E+\bar{E}) U\D_\al U+i(\iota_{\bar{\mcW}})_{\dot\al}( E+\bar{E}) U\bar{\D}^{\dot\al}U, \\
g_4 &= 3E^2G+3EW^\al W_\al-\frac{i}{2}\bar{\D}^2(( 2E+\bar{E}) UH+2U\D^\al UW_\al+U\bar{\D}_{\dot\al}U\bar{W}^{\dot\al}\non\\
& \qquad -iU\D^\al U(\iota_\mcW)_\al U+iU\bar{\D}_{\dot\al}U(\iota_{\bar{\mcW}})^{\dot\al}U).
\end{align}
\end{subequations}
This result is new and would be relevant, for example, for a reduction of seven-dimensional supergravity to four dimensions.

\section{Conclusions}
\label{S:Conclusions}

In this paper we have elucidated the supergeometry underlying gauged $p$-form hierarchies in 4D, $N=1$ superspace and used it to construct all of the associated Chern-Simons-like invariants.
This was done by describing the four-dimensional part in terms of superforms and extending the resulting de Rham complex to a double complex of forms covariantly coupled to non-abelian gauge fields.
This construction defines field-strengths of the super-de Rham complex which can be wedged together to make superforms that extend the bosonic Chern-Simons forms.
For each such Chern-Simons superform, we could construct a second manifestly gauge-covariant superform such that their difference was closed (Weil triviality).
These closed forms then defined manifestly supersymmetric actions by the ectoplasm method of integration.

For the cubic Chern-Simons invariant, this procedure recovers the action first constructed in \cite{Becker:2016rku} and used in \cite{Becker:2016edk} to compute the exact scalar potential of M-theory on backgrounds with $G_2$ structure.
More generally, these constructions are expected to apply to higher-dimensional/extended supergravity theories.
As an illustration of our general solution, we explicitly wrote out the quartic invariant which is new and could, in principle, be used to compute the scalar potential for seven-dimensional supergravity.

Extensions of the method to $N=2$ superspace \cite{Butter:2011sr}, 5D, $N=1$ \cite{Butter:2014xxa}, and 6D, $N=(1,0)$ \cite{Linch:2012zh, Arias:2014ona, Butter:2016qkx, Butter:2017jqu} should also be possible. 
Complications arise in these superspaces in (at least) two ways.
First, if we intend to keep the representations off-shell, we must consider embeddings into projective \cite{Karlhede:1984vr, Lindstrom:1987ks, Lindstrom:1989ne} and harmonic \cite{Galperin:1984av, Galperin:2001uw} superspaces \cite{Siegel:1981dx}.
Second, there is an additional subtlety if we wish to connect super-$p$-forms to their bosonic counterparts at the component level (although this is not necessary for the abstract construction). It is well-known that the 4D, $N = 1$ super-de Rham complex of irreducible super-$p$-forms defines multiplets which include bosonic $p$-forms. In superspaces with more than four supercharges this is no longer universally true. In 5D, $N = 1$ superspace, for example, the irreducible ``3-form" is instead a multiplet of superconformal gauge parameters \cite{Gates:2014cqa, Linch:2014iza}.

Eventually, one would like to go beyond the computation of scalar potentials and obtain the effective action for such theories complete with gravity couplings (including all gravitino superfields).
This is not trivial and it would be interesting to know what the conditions on the general hierarchy might be that would make this possible. 
Partial results come from minimal coupling to old-minimal supergravity \cite{Randall:2016zpo} or, as emphasized in \cite{Aoki:2016rfz, Yokokura:2016xcf}, even more simply from conformal superspace \cite{Butter:2009cp, Butter:2010sc}.

\section*{Acknowledgements}
W{\sc dl}3 and D{\sc r} are grateful to the Simons Center for Geometry and Physics for hospitality during the {\sc ix} Simons Summer Workshop. S{\sc r} is similarly thankful to the Mitchell Institute for hosting in-person collaboration between the authors.
K{\sc b}, M{\sc b}, W{\sc dl}3, and D{\sc r} further thank the Simons Center for Geometry and Physics for generous financial and logistical support of the Simons Workshop {\it String Theory and Scattering Amplitudes}.
This work is partially supported by NSF Focused Research Grant DMS-1159404 and the Mitchell Institute for Physics and Astronomy at Texas A\&M University.

\appendix
\section{Superform Components}
\label{S:MoreForms}

In section \ref{S:PrepotentialNATH}, we reviewed the embedding of the bosonic $p$-form potentials into superfield prepotentials (cf.\ eq.\ \ref{subeqns:FormEmbed}). In this appendix, we further embed these components and prepotentials in superforms. It is not necessary to understand the details of this embedding to construct the Chern-Simons super-invariants, but we will be explicit in our presentation as they are needed to project the superspace action to components and because we use some of the expressions here in the proofs of section \ref{S:AllCS}.

We begin with the zero-form bosonic potential, the axion, which was given by
\be
a=\hlf(\Phi+\bar{\Phi})\big|.
\ee
It is clear how we can lift this to a super-zero-form, also known as a superfield; we just remove the $\big|$ and write
\be
C^{(0)}=\hlf(\Phi+\bar{\Phi}).
\ee

In this section our goal is to find superforms $C^{(p)}$ which will have the bosonic potentials sitting in their bottom components and a set of field-strengths $F^{(p)}$ related to them by $F^{(p+1)}=dC^{(p)}$. Since we have already defined $C^{(0)}$, we can compute
\be
F^{(1)}_s=\hlf D_s\Phi,\quad F^{(1)}_{\bar{s}}=\hlf\bar{D}_{\bar{s}}\bar{\Phi},\quad F^{(1)}_\psi=\hlf\p_\psi(\Phi+\bar{\Phi}).
\ee
With some foresight, we can rewrite these expressions in terms of the superfield $U=(\Phi-\bar{\Phi})/2i$,
\be
F^{(1)}_s=iD_sU,\quad F^{(1)}_{\bar{s}}=-i\bar{D}_{\bar{s}}U,\quad F^{(1)}_\psi=-\frac{1}{4}\bar{\s}_\psi^{\dot\al\al}[ D_\al,\bar{D}_{\dot\al}] U.
\ee
Note the similarity between $F^{(1)}_\psi$ and (\ref{subeqn:AEmbed}).  We can take this as a sign to make the (justifiable \cite{Randall:2016zpo, Linch:2014iza}) ansatz that
\be
C^{(1)}_s=iD_sV,\quad C^{(1)}_{\bar{s}}=-i\bar{D}_{\bar{s}}V,\quad C^{(1)}_\psi=-\frac{1}{4}\bar{\s}_\psi^{\dot\al\al}[ D_\al,\bar{D}_{\dot\al}] V.
\ee
That is, we have simply replaced $U$ with $V$ to go from $F^{(1)}$ to $C^{(1)}$.  But now that we have $C^{(1)}$, we can compute $F^{(2)}=dC^{(1)}$,
\be
F^{(2)}_{ss}=F^{(2)}_{s\bar{s}}=F^{(2)}_{\bar{s}\bar{s}}=0,\quad F^{(2)}_{s\psi}=-(\s_\psi)_{s\dot\al}\bar{W}^{\dot\al},\quad F^{(2)}_{\bar{s}\psi}=-(\s_\psi)_{\al\bar{s}}W^\al,\non
\ee
\be
F^{(2)}_{\psi\psi}=-\frac{i}{2}[(\s_{\psi\psi})_\al^{\hph{\al}\beta}D^\al W_\beta-(\bar{\s}_{\psi\psi})^{\dot\al}_{\hph{\dot\al}\dot\beta}\bar{D}_{\dot\al}\bar{W}^{\dot\beta}].
\ee
Using the same trick we substitute $\Sigma^\al$ in for $W^\al$ to get $C^{(2)}$,
\be
C^{(2)}_{ss}=C^{(2)}_{s\bar{s}}=C^{(2)}_{\bar{s}\bar{s}}=0,\quad C^{(2)}_{s\psi}=-(\s_\psi)_{s\dot\al}\bar{\Sigma}^{\dot\al},\quad C^{(2)}_{\bar{s}\psi}=-(\s_\psi)_{\al\bar{s}}\Sigma^\al,\non
\ee
\be
C^{(2)}_{\psi\psi}=-\frac{i}{2}[(\s_{\psi\psi})_\al^{\hph{\al}\beta}D^\al\Sigma_\beta-(\bar{\s}_{\psi\psi})^{\dot\al}_{\hph{\dot\al}\dot\beta}\bar{D}_{\dot\al}\bar{\Sigma}^{\dot\beta}].
\ee
Then $F^{(3)}=dC^{(2)}$ has non-vanishing components
\be
F^{(3)}_{s\bar{s}\psi}=i(\s_\psi)_{s\bar{s}}H,\quad F^{(3)}_{s\psi\psi}=(\s_{\psi\psi})_s^{\hph{s}\al}D_\al H,\quad F^{(3)}_{\bar{s}\psi\psi}=-(\bar{\s}_{\psi\psi})^{\dot\al}_{\hph{\dot\al}\bar{s}}\bar{D}_{\dot\al}H,\non
\ee
\be
F^{(3)}_{\psi\psi\psi}=\frac{1}{8}\e_{\psi\psi\psi a}(\bar{\s}^a)^{\dot\al\al}[ D_\al,\bar{D}_{\dot\al}] H.
\ee
Similarly,
\be
C^{(3)}_{s\bar{s}\psi}=i(\s_\psi)_{s\bar{s}}X,\quad C^{(3)}_{s\psi\psi}=(\s_{\psi\psi})_s^{\hph{s}\al}D_\al X,\quad C^{(3)}_{\bar{s}\psi\psi}=-(\bar{\s}_{\psi\psi})^{\dot\al}_{\hph{\dot\al}\bar{s}}\bar{D}_{\dot\al}X,\non
\ee
\be
C^{(3)}_{\psi\psi\psi}=\frac{1}{8}\e_{\psi\psi\psi a}(\bar{\s}^a)^{\dot\al\al}[ D_\al,\bar{D}_{\dot\al}] X,
\ee
\be
F^{(4)}_{ss\psi\psi}=4(\s_{\psi\psi}\e)_{ss}\bar{G},\quad F^{(4)}_{\bar{s}\bar{s}\psi\psi}=-4(\e\bar{\s}_{\psi\psi})_{\bar{s}\bar{s}}G,\quad F^{(4)}_{s\psi\psi\psi}=\hlf\e_{\psi\psi\psi a}\s^a_{s\dot\al}\bar{D}^{\dot\al}\bar{G},\non
\ee
\be
F^{(4)}_{\bar{s}\psi\psi\psi}=\hlf\e_{\psi\psi\psi a}\s^a_{\al\bar{s}}D^\al G,\quad F^{(4)}_{\psi\psi\psi\psi}=\frac{i}{8}\e_{\psi\psi\psi\psi}( D^2G-\bar{D}^2\bar{G}),
\ee
and
\be
C^{(4)}_{ss\psi\psi}=4(\s_{\psi\psi}\e)_{ss}\bar{\G},\quad C^{(4)}_{\bar{s}\bar{s}\psi\psi}=-4(\e\bar{\s}_{\psi\psi})_{\bar{s}\bar{s}}\G,\quad C^{(4)}_{s\psi\psi\psi}=\hlf\e_{\psi\psi\psi a}\s^a_{s\dot\al}\bar{D}^{\dot\al}\bar{\G},\non
\ee
\be
C^{(4)}_{\bar{s}\psi\psi\psi}=\hlf\e_{\psi\psi\psi a}\s^a_{\al\bar{s}}D^\al\G,\quad C^{(4)}_{\psi\psi\psi\psi}=\frac{i}{8}\e_{\psi\psi\psi\psi}( D^2\G-\bar{D}^2\bar{\G}).
\ee
One can check that $dC^{(4)}=0$, so the hierarchy stops here.

Moreover, the gauge transformations can also be cast in this language with
\be
\d C^{(p)}=d\La^{(p-1)},
\ee
where the components of the forms $\La^{(p-1)}$ are obtained by taking the expressions for $C^{(p-1)}$ and substituting $\La$, $L$, $\Upsilon^\al$, and $\Xi$ for $\Phi$, $V$, $\Sigma^\al$, and $X$ respectively.  The field-strengths $F^{(p+1)}$ are invariant by the nilpotency of the super-de Rham differential, $\d F^{(p+1)}=d^2\La^{(p-1)}=0$.

For the non-abelian tensor hierarchy, we can almost take the same expressions for the polyforms $C$ and $F=\Q C$ as in the abelian tensor hierarchy \cite{Becker:2016xgv} but using the full non-abelian expressions (\ref{eqs:SuperfieldStrengths}). The only additional modifications we need to make are in degree two, where we need to modify the expressions as
\be
C^{(2)}_{s\psi}=-(\s_\psi)_{s\dot\al}(\bar{\Sigma}^{\dot\al}+i(\iota_{\bar{\mcW}})^{\dot\al}V),\quad C^{(2)}_{\bar{s}\psi}=-(\s_\psi)_{\al\bar{s}}(\Sigma^\al-i(\iota_\mcW)^\al V),\non
\ee
\begin{align}
C^{(2)}_{\psi\psi}& =(\s_{\psi\psi})_\al^{\hph{\al}\beta}\left( -\frac{i}{2}\D^\al\Sigma_\beta-\hlf(\iota_{\D\mcW})^\al_{\hph{\al}\beta}V+(\iota_\mcW)_\beta\D^\al V\right)\\
& \qquad +(\bar{\s}_{\psi\psi})^{\dot\al}_{\hph{\dot\al}\dot\beta}\left(\frac{i}{2}\bar{\D}_{\dot\al}\bar{\Sigma}^{\dot\beta}-\hlf(\iota_{\bar{\D}\bar{\mcW}})_{\dot\al}^{\hph{\dot\al}\dot\beta}V+(\iota_{\bar{\mcW}})^{\dot\beta}\bar{\D}_{\dot\al}V\right)\!,
\end{align}
and similarly for $F^{(2)}$ with $\Sigma \rightarrow W$ and $V \rightarrow U$. For a more explicit step-by-step derivation, we refer to \cite{Randall:2016zpo}.

As before, the gauge symmetry is simply $\d C=\Q\La$ with $\d F=\Q^2\La=0$ by the nilpotency of $\Q$.  The expressions for $\La$ are obtained by substituting the gauge parameter superfields in for the prepotentials in the expressions for $C$ (including the given modifications in degree two).

\subsection{Composite Forms}

We also need to be able to take wedge products of superforms.  For the four-dimensional part of the forms, this works in a straightforward way with numerical factors determined in the same way as for ordinary differential forms and signs determined by the character of the relevant indices (\textit{i.e.}, spinor or vector). The wedge product of two polyforms is given by extending the usual wedge product by linearity.  For our purposes there is one additional wrinkle, which is that our objects are also ordinary differential forms in the internal space. Since we generally write our objects with the spacetime indices explicit and the internal indices implicit, the wedge product of two objects $\om\w\xi$ can get an additional sign when the internal degree of $\om$ and the spacetime degree of $\xi$ are both odd.  As examples, here are the rules for constructing the components of $F\w F$ up to degree two\footnote{Recall that the total (internal + spacetime) degree of $F$ is four, so the parity of the spacetime degree and the internal degree is the same.}
\begin{subequations}
\begin{align}
( F\w F)^{(0)} &= F^{(0)}F^{(0)},\\
( F\w F)^{(1)}_s &= 2F^{(0)}F^{(1)}_s,\\
( F\w F)^{(1)}_\psi &= 2F^{(0)}F^{(1)}_\psi,\\
( F\w F)^{(2)}_{ss} &= 2F^{(0)}F^{(2)}_{ss}-2F^{(1)}_sF^{(1)}_s,\\
( F\w F)^{(2)}_{s\bar{s}} &= 2F^{(0)}F^{(2)}_{s\bar{s}}-2F^{(1)}_sF^{(1)}_{\bar{s}},\\
( F\w F)^{(2)}_{s\psi} &= 2F^{(0)}F^{(2)}_{s\psi}-2F^{(1)}_sF^{(1)}_\psi,\\
( F\w F)^{(2)}_{\psi\psi} &= 2F^{(0)}F^{(2)}_{\psi\psi}-2F^{(1)}_\psi F^{(1)}_\psi.
\end{align}
\end{subequations}
Of course some of these are zero since $F^{(2)}_{ss}=F^{(2)}_{s\bar{s}}=0$, but those terms are included to give the general pattern.  For a more complicated relevant example,
\be
( F\w F)^{(4)}_{\bar{s}\bar{s}\psi\psi}=2F^{(0)}F^{(4)}_{\bar{s}\bar{s}\psi\psi}-4F^{(1)}_{\bar{s}}F^{(3)}_{\bar{s}\psi\psi}-4F^{(1)}_\psi F^{(3)}_{\bar{s}\bar{s}\psi}+2F^{(2)}_{\bar{s}\bar{s}}F^{(2)}_{\psi\psi}-8F^{(2)}_{\bar{s}\psi}F^{(2)}_{\bar{s}\psi}.
\ee

\renewcommand*{\refname}{\vspace*{-1em}}


\end{document}